\documentclass[review]{elsarticle}

\usepackage{lineno,hyperref}
\modulolinenumbers[5]
\usepackage{epsfig} 
\usepackage{graphicx}
\usepackage{caption}
\usepackage{subcaption}
\usepackage{multirow}
\usepackage{multicol}
\usepackage{etoolbox}
\usepackage{setspace}
\usepackage{lipsum}
\makeatletter
\def\ps@pprintTitle{%
 \let\@oddhead\@empty
 \let\@evenhead\@empty
 \def\@oddfoot{}%
 \let\@evenfoot\@oddfoot}
\makeatother
\usepackage{amssymb,mathtools}
\usepackage{lscape}

\usepackage[utf8]{inputenc}

\usepackage{siunitx}
\usepackage[nocfg]{nomencl}
\usepackage{nomencl}
\usepackage{ifthen}
\renewcommand\nomgroup[1]{%
  \ifthenelse{\equal{#1}{A}}{%
    \item[\textbf{Acronyms}]}{
  \ifthenelse{\equal{#1}{R}}{%
    \item[\textbf{Roman Symbols}]}{
  \ifthenelse{\equal{#1}{G}}{%
    \item[\textbf{Greek Symbols}]}{
  \ifthenelse{\equal{#1}{S}}{%
    \item[\textbf{Superscripts}]}{
  \ifthenelse{\equal{#1}{U}}{%
    \item[\textbf{Subscripts}]}{
  \ifthenelse{\equal{#1}{X}}{%
    \item[\textbf{Other Symbols}]}{
  {}}}}}}}}

\makenomenclature

\usepackage{fancyhdr}
\date{}
\title{Effects of Diffuse and Collimated Beam Radiation on a Symmetrical Cooling Case of Natural Convection}

\author{G. Chanakya, Pradeep Kumar \\
 Numerical Experiment Laboratory \\
(Radiation \& Fluid Flow Physics)\\ 
School of Engineering\\ 
Indian Institute of Technology Mandi \\
Mandi 175005, India.\\
Email: pradeepkumar@iitmandi.ac.in
}
\begin{document}

\maketitle

\section*{Abstract}
In the present work, the effects of diffuse and collimated radiation on the symmetrical cooling case of natural convection in a two-dimensional cavity heated from the bottom have been investigated, numerically. The collimated beam radiation feature has been developed in OpenFOAM framework and other libraries of fluid flow and heat transfer have been combined together for a coupled simulation of fluid flow and heat transfer. The cavity is convectively heated from the bottom with heat transfer coefficient of 50 $W/m^2$K and free stream temperature 305 K, while both vertical walls of cavity are isothermal at temperature of 296 K. The top wall is adiabatic and all walls are opaque for radiation heat transfer. For collimated case, a small semitransparent window of non-dimensional width of 0.05 at height of 0.7 has been created on the left wall and a collimated irradiation of value 1000 $W/m^2$ at an angle of $45^0$ is applied on this semitransparent window. The study has been performed in two stages, first, the effect of diffuse radiation on the natural convection has been observed and then collimated beam is passed through the semitransparent window. The results reveal that the diffuse radiation has little effect on the dynamics of two rolls inside the cavity, however, collimated beam irradiation changes the dynamics of two rolls significantly and also the heat transfer characteristics. This further changes with the optical thickness of the fluid. The left vortex is bigger in size than the right vortex for collimated beam in transparent fluid, whereas, reverse trend is seen for collimated beam in non-zero optical thickness of the fluid. The size of the left vortex increases with the increase of optical thickness of the fluid. The heat transfer reversal happens at the zone of collimated beam incident on the bottom wall for collimated beam in transparent medium, whereas, this does not happen for participating medium. 

\section*{keywords}
Semitransparent Window; Natural Convection; Collimated Beam; Symmetrical Cooling; Irradiation; Bottom Heating;

\section{Introduction}

The study of natural convection in a square/rectangular cavities combined with radiation provides the fundamental insight of fluid flow and heat transfer for many engineering applications. This is the prime reason for multi-physics problems in a simple domain that became interest to researchers from last few decades.

The natural convection in a square/cylindrical enclosures with heating from the bottom and cooled from the sides, with different inclination angles were analysed by many researchers \cite{Torrance,Calcagani,Ganzorolli,Aydin} and the formation of two symmetrical vortices were observed. While, the single vortex formation has been observed in the inclined cavity case \cite{Acharya}.

Webb and Viskanta \cite{Webb} experimentally studied the radiation induced buoyancy flow in a rectangular enclosure which was irradiated from a side. They have observed the formation of thin thermal boundary layers along the vertical walls where the flow structure lost the centrosymmetry characteristics. They also developed a theoretical model for the prediction of radiative heating of the fluid and motion induced due to buoyancy. Mezrhab et al. \cite{Mezrhab} and Sun et al. \cite{Hua} had numerically studied the combined natural convection and radiation for a centrally located hot square body in a cavity and observed that the radiation homogenized the temperature inside the cavity, whereas emissivity of the surfaces have affected the fluid flow characteristics inside the cavity. This was due to strengthening of recirculating zones which resultant into the stabilizing of flow fields due to surface radiation exchange. Mukul et al. \cite{Mukul} have studied the critical assessment of interaction of thermal radiation with natural convection in a square enclosure with different geometric configurations using three different approaches, like, incompressible (with Boussinesq approximation), purely compressible and quasi-incompressible (i.e low Mach-number approximation) approach. It was observed that the presence of corners had affected the heat transfer locally. Kumar and Eswaran \cite{Kumar} investigated the effect of radiation on natural convection in two slanted cavities of angles $45^0$ and $60^0$ and reported that radiation effect was more pronounced for cavity with slanted angle $60^0$. Saravanan and Sivaraj \cite{Saravanan} had reported the results of natural convection with thermal radiation for a non-uniformly heated thin plate placed at horizontal and vertical directions in a cavity. They noticed that the overall heat transfer had increased for the horizontal placement of this plate while it has decreased for the vertical placement.

The natural convection with the volumetric radiation in a square cavity was studied by Mondal and Mishra \cite{Bittagopal} where they used lattice Boltzmann method for the fluid flow and finite volume method for radiative transfer equation (RTE). It was observed that scattering coefficient had no significant effect on the stream lines while extinction coefficient had a significant effect on the isothermal lines distribution.

The performance of DOM, FVM, P1, SP3 and P3 methods for RTE in a two-dimensional absorbing/emitting medium were studied by Sun et al. \cite{Yujia}. The Monte Carlo method had been used as a benchmark solution for comparison of performance of above different method of RTE. Xing et al. \cite{Yuan} have studied the natural convection with heated circular, elliptical, square and triangular geometries inside cylindrical enclosure. The effect of surface radiation was incorporated and concluded that the presence of corners and larger upper space had a major effect on the heat transfer performance.

In all the above works diffuse radiation was considered, however, little work is available on collimated beam radiation, like, work by Anand and Mishra \cite{Anand} and Ben and Dez \cite{Ben}, both the work had tried to capture the collimated beam bending phenomenon in a graded refractive index medium by discrete ordinate method for RTE. The numerical study of a collimated beam in the varying refractive index medium was studied by Ilyushin \cite{Ya}. Few authors have also studied the short pulse collimated irradiation phenomenon \cite{Anil,Rath}.

From the above literatures, it has been observed that there is not much work performed for the effects of collimated beam radiation on natural convection. The collimated beam radiation has numerous applications like solar heating of room, solar cavity receiver in solar thermal power plants, optics etc. The collimated beam radiation feature had been developed and integrated with libraries of other fluid flow and heat transfer models in OpenFOAM \cite{openfoam2017open} framework and an application had also been developed to simulate natural convection with collimated beam irradiation.

In the present work, the effect of diffuse and collimated beam radiation has been investigated in symmetrical cooling case of natural convection in a two-dimensional cavity heated at the bottom. The manuscript is organized as follows: The next section deals with the problem statement and mathematical modeling and its numerical procedure has been described in section 3. The results are mostly explained in non-dimensional numbers and hence these non-dimensional numbers are expressed in section 3.1, followed by validation in section 4 and grid independent test study in section 5. The section 6 elaborates results and discussion and finally the present work is concluded in section 7.

\section{Problem Statement}
\begin{figure}[b]
 \centering\includegraphics[width=6cm]{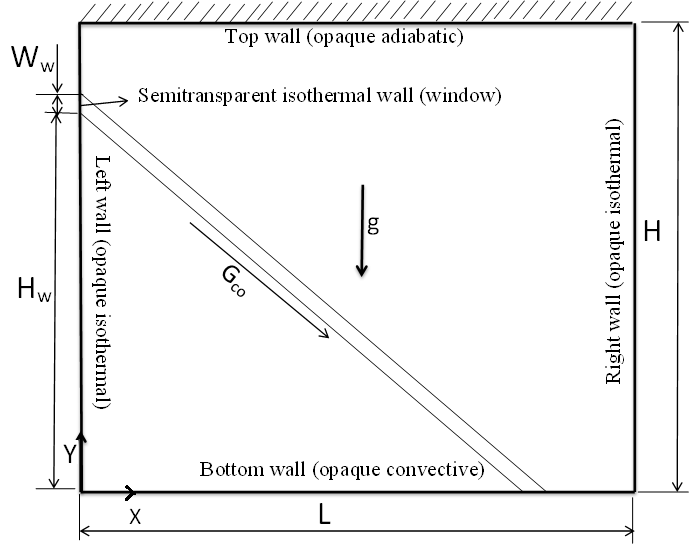}
\caption{Schematic diagram for the study of collimated irradiation in a cavity}
\label{prob_def}
\end{figure}
Figure \ref{prob_def} depicts a square cavity (L=H) which has a semitransparent window of a non-dimensional width 0.05 at non-dimensional height of 0.7 on the left wall. Both the vertical walls as well as semitransparent window are isothermal at temperature of 296 K, whereas bottom wall is convectively heated with free stream temperature of 305K and heat transfer coefficient of 50 $W/m^2$K. The top wall is subjected to adiabatic boundary and all four walls of the cavity are opaque to the radiation. The window is treated as opaque for diffuse radiation analysis whereas semitransparent for collimated radiation. Collimated beam has been passed through semitransparent window at an angle of $45^0$. Natural convection is established inside the cavity due to temperature difference. The buoyancy force inside the cavity corresponds to Rayleigh number $10^5$ based on the temperature difference between free stream and vertical wall temperature and the Prandtl number of fluid is 0.71. The Euclidean axis are along the horizontal and the vertical walls of cavity and origin is at lower left corner of the cavity. The gravity force act in the negative Y direction.

\section{Mathematical modelling and Numerical procedures}
The following assumptions have been considered for the mathematical modelling of the problem described below:
\begin{enumerate}
    \item Flow is steady, laminar, incompressible and two dimensional.
    \item Flow is driven by buoyancy force that is modeled by Boussinesq approximation.
    \item The thermophysical properties of fluid is constant.
    \item The fluid may or may not participate in radiative heat transfer.
    \item The refractive index of the medium is constant and equal to one.
    \item The fluid absorbs and emits but does not scatter the radiation energy.
\end{enumerate}
Based on the above assumptions the governing equations in the Cartesian coordinate system are given by  

\begin{equation} \label{mass:equN} 
\frac{\partial u_i}{\partial x_i} = 0  
\end{equation}
\vspace{-1cm}
\begin{equation} \label{momentum}
\frac{\partial  u_i u_j}{\partial x_j}=-\frac{1}{\rho}\frac{\partial p}{\partial x_i} + \nu\frac{\partial^2u_i}{\partial x_j\partial x_j}+g \beta_{T}(T-T_{c})\delta_{i2}  
\end{equation}

\begin{equation}\label{energy}
\frac{\partial u_jT}{\partial x_j} = \frac{k}{\rho C_p}\frac{\partial^2T}{\partial x_j\partial x_j} - \frac{1}{\rho C_p} \frac{\partial q_{R}}{\partial x_i} 
\end{equation}
\vspace{2cm}
\noindent where $\frac{\partial q_{R_{i}}}{\partial x_{i}}$ is the divergence of the radiative flux, which can be calculated as
\vspace{-1cm}
\begin{eqnarray}
\frac{\partial q_{R_{i}}}{\partial x_{i}}=\kappa_{a}(4\pi I_b-G)\label{div_eq}
\end{eqnarray}
where $\kappa_{a}$ is the absorption coefficient, $I_b$ is the black body intensity and $G$ is the irradiation which is evaluated by integrating the radiative intensity ($I$) in all directions, i.e.,
\begin{eqnarray}
G=\int_{4\pi} I d\Omega
\end{eqnarray}
The intensity field inside the cavity can be obtained by solving the following radiative transfer equation (RTE) 
 
\begin{equation} 
\label{equN:radiation_1}
\frac{\partial I(\bf \hat{r},\bf \hat{s})}{\partial s}=\kappa_{a}I_{b}(\bf\hat{s})-(\kappa_{a}) \text{I} (\bf\hat{r},\bf \hat{s})
\end{equation}
Where $\bf \hat{r},\bf \hat{s}$ is position and direction vectors. Whereas s is path length. 
The Navier-Stokes equation and temperature equations are subjected to following boundary conditions 
\begin{enumerate}
\centering
 \item[]\textit{Flow boundary condition}
    \item[] Cavity walls: u=v=0
 \item[] \textit{Thermal boundary conditions}
 
    \item[1] Left wall at x=0;\hspace{0.2cm} T= 296K
    \item[2] Right wall at x=1;\hspace{0.2cm} T=296K
    \item[3] Bottom wall y=0 ;\hspace{0.2cm} $q_{conv}= h_{free} (T_{free}-T_{w})$
    \item[4] Top wall at y=1;\hspace{0.2cm} $q_{c}+q_{r}=0$
    \item[]where $q_c=-k \frac{\partial T}{\partial n}$ and $ q_{r}=\int_{4\pi}I({\bf r_w})({\bf \hat{n}\cdot\hat{s}})\mathrm{d}\Omega $
\end{enumerate}

The radiative transfer equation (\ref{equN:radiation_1}) is subjected to following boundary condition, all cavity wall (assumed black wall) except semitransparent window 

\begin{eqnarray}
\noindent I({\bf r_w,\hat{s}})=\epsilon_w I_b({\bf r_w})+\frac{1-\epsilon_w}{\pi}\int_{\bf \hat{n}\cdot\hat{s}>0}I({\bf r_w,\hat{s}})|{\bf \hat{n}\cdot\hat{s}}|\mathrm{d}\Omega.\nonumber\\
\mbox{for}~~{\bf \hat{n}\cdot\hat{s}}<0 \label{rte:bound2}
\end{eqnarray}

where $\hat{n}$ is the surface normal and the emissivity of all walls is considered to be 1.

The semitransparent window is subjected to collimated irradiation $(G_{co})$ of value 1000 $W/m^2$. The boundary condition for RTE on semitransparent window is 

\begin{eqnarray}
\noindent I({\bf r_w,\hat{s}})=I_{co}({\bf r_w,\hat{s}}) \delta (\theta-45^{0})+\epsilon_w I_b({\bf r_w})+\frac{1-\epsilon_w}{\pi}\int_{\bf \hat{n}\cdot\hat{s}>0}I({\bf r_w,\hat{s}})|{\bf \hat{n}\cdot\hat{s}}|\mathrm{d}\Omega.\nonumber\\
\mbox{for}~~{\bf \hat{n}\cdot\hat{s}}<0 \label{rte:bound3}
\end{eqnarray}
where $\delta (\theta-45^{0})$ is Dirac-delta function,
\begin{equation}
  \delta(\theta-45^{0})=
    \begin{cases}
            1, &         \text{if } \theta=45^{0},\\
            0, &         \text{if } \theta \neq 45^{0}.
    \end{cases}  
\end{equation}

\begin{figure}[t]
 \begin{subfigure}{4cm}
    \centering\includegraphics[width=5cm]{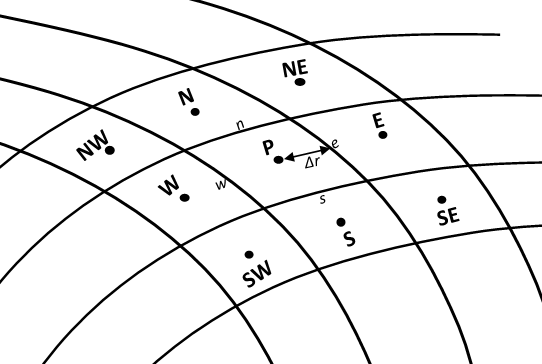}
    \caption{}
    \label{2d_fvm}
  \end{subfigure}\hfill
   \begin{subfigure}{5cm}
    \centering\includegraphics[width=6cm]{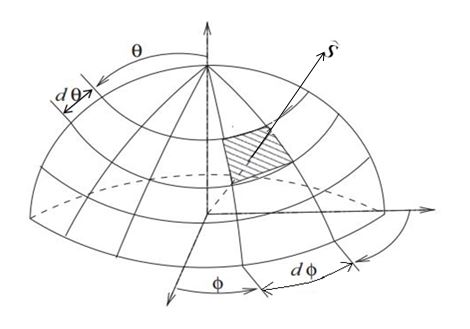}
    \caption{ }
    \label{angular}
  \end{subfigure}\hfill
  \caption{Pictorial representation of (a) typical volumetric cell in two dimensional case for finite volume method and (b) Virtual angular discreatization to obtain directions for radiative transfer equation }
\label{grid_space_ang}
\end{figure}

\begin{figure}[!htb]
 \begin{subfigure}{4cm}
    \centering\includegraphics[width=5.5cm]{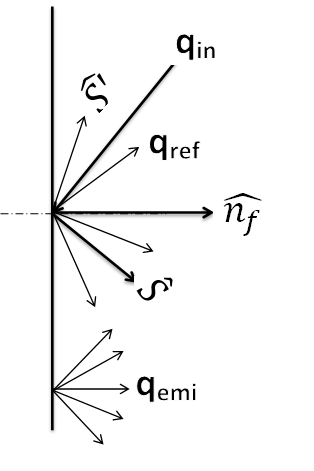}
    \caption{}
    \label{Diff_BC}
  \end{subfigure}\hfill
   \begin{subfigure}{5cm}
    \centering\includegraphics[width=5cm]{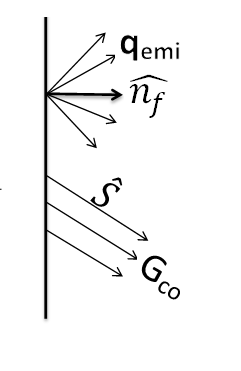}
    \caption{}
    \label{semi_colli}
  \end{subfigure}\hfill
  \caption{Pictorial representation of (a) Diffuse reflection of a incident ray and diffuse emission due to wall temperature on opaque wall (b) Diffuse emission and collimated transmission from a semitransparent window}
\label{Diff_ref_wall}
\end{figure}

$I_{co}$ is intensity of collimated irradiation and calculated from the irradiation value as below
\begin{equation}
    I_{co}=\frac{G_{co}}{\mathrm{d}\Omega}
\end{equation}
Where $d\Omega$ is the collimated beam width. In the current work, the solid angle of discretized angular space (Fig. \ref{angular}) in collimated direction is considered as beam width of the collimated beam. The pictorial representation of  diffuse emission and reflection and collimated beam radiation from the wall is shown in Fig. \ref{Diff_ref_wall}. The collimated feature has been developed in OpenFOAM framework, an open source software and coupled with other fluid and heat transfer libraries. The combined application has been used for numerical simulation. The OpenFOAM uses the finite volume method (FVM) to solve the Navier-Stokes and the energy equations. The FVM integrates an equation over a control volume (Fig. \ref{2d_fvm}) to convert the partial differential equation into a set of algebraic equations in the foam
\begin{equation}
    a_{p}\phi_{p}= \sum_{nb}a_{nb}\phi_{nb}+S
\end{equation}
where $\phi_{p}$ is any scalar and $a_{p}$ is central coefficient, $a_{nb}$ coefficients of neighbouring cells and S is the source values, whereas, RTE eq (\ref{equN:radiation_1}) is converted into a set of algebraic equations by double integration over a control volume and over a control angle. The set of algebraic equations are solved by Preconditioned bi-conjugate gradient (PBiCG) and the details of the algorithm can be found in the book by Patankar \cite{patankar} and Moukalled \cite{Moukalled}. In the present simulation, linear upwind scheme which is second order accurate has been used to interpolate face centred value. The linear upwind scheme is given mathematically as
\begin{equation}
   \phi_{f}=
    \begin{cases}
            \phi_{p}+ \nabla\phi \cdot \nabla r, &         \text{if} f_{\phi} > 0,\\
            \phi_{nb}+ \nabla\phi \cdot \nabla r, &         \text{if} f_{\phi} < 0.
    \end{cases}   
\end{equation}
and $f_{\phi}$ is the flux of the scalar $\phi$ on a face (Fig \ref{2d_fvm}).

 \subsection{Non-dimensional Parameters}
The OpenFOAM simulation produces the results in dimensional quantities. To explain the results in more general form, the simulated results are converted into non-dimensional parameters. The scales for length, velocity, temperature, and conductive and radiative fluxes are L, u$_o$, (T$_{free}$-T$_{c}$), $\kappa$(T$_{free}$-T$_{c}$)/L and $\sigma$T$_{free}^{4}$ respectively, where $u_{o}=\sqrt {L g \beta (T_{free}-T_{c})}$ is convective velocity scale.

The non-dimensional quantities and parameters involved in the present problem are as follows,
\begin{eqnarray*}
U =\frac{u}{u_{o}}  \hspace{0.5cm}   V=\frac{v}{u_{o}} \hspace{0.5cm}  \hspace{0.5cm} X =\frac{x}{L}  \hspace{0.5cm}   Y=\frac{y}{L} \hspace{0.5cm} \theta =\frac{T-T_{c}}{T_{free}-T_{c}}   
\end{eqnarray*}

\begin{eqnarray*}
Ra=\frac{g \beta (T_{free}-T_{c})L^{3}}{\nu \alpha} \hspace{0.5cm}  Pr = \frac{\nu}{\alpha} 
\end{eqnarray*}

The optical thickness is defined as $\tau=\kappa_{a} L$ and non-dimensional irradiation is given as
\begin{eqnarray*}
\overline{G}=\frac{G}{\sigma T^{4}_{free}} \hspace{1cm}
\end{eqnarray*}

The Nu$_{C}$ and Nu$_{R}$ are conductive and radiative Nusselt numbers respectively and defined as
\begin{eqnarray*}
Nu_{C}=\frac{q_{Cw}L}{k(T_{free}-T_{c})} \hspace{1cm} Nu_{R}=\frac{q_{Rw}L}{k(T_{free}-T_{c})} \hspace{1cm} 
\end{eqnarray*}
Thus, the total Nusselt number is defined as below,
\begin{eqnarray*}
 Nu=Nu_{C}+Nu_{R} \hspace{0.5cm}  \hspace{0.5cm}  
\end{eqnarray*}

\section{Validation}

\begin{figure}[t]
 \begin{subfigure}{5cm}
    \centering \includegraphics[width=8cm]{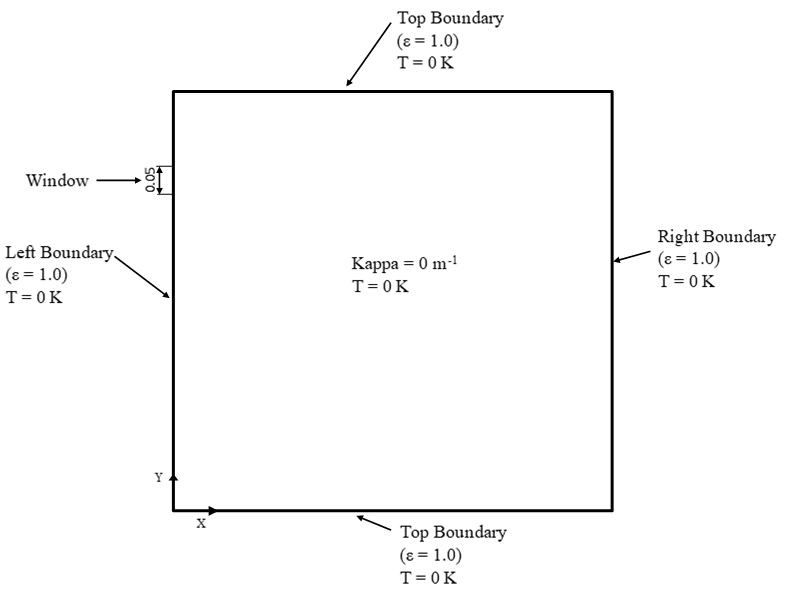}
    \caption{}
    \label{collimated_geo}
  \end{subfigure}\hfill
   \begin{subfigure}{5cm}
    \centering
    \includegraphics[width=6cm]{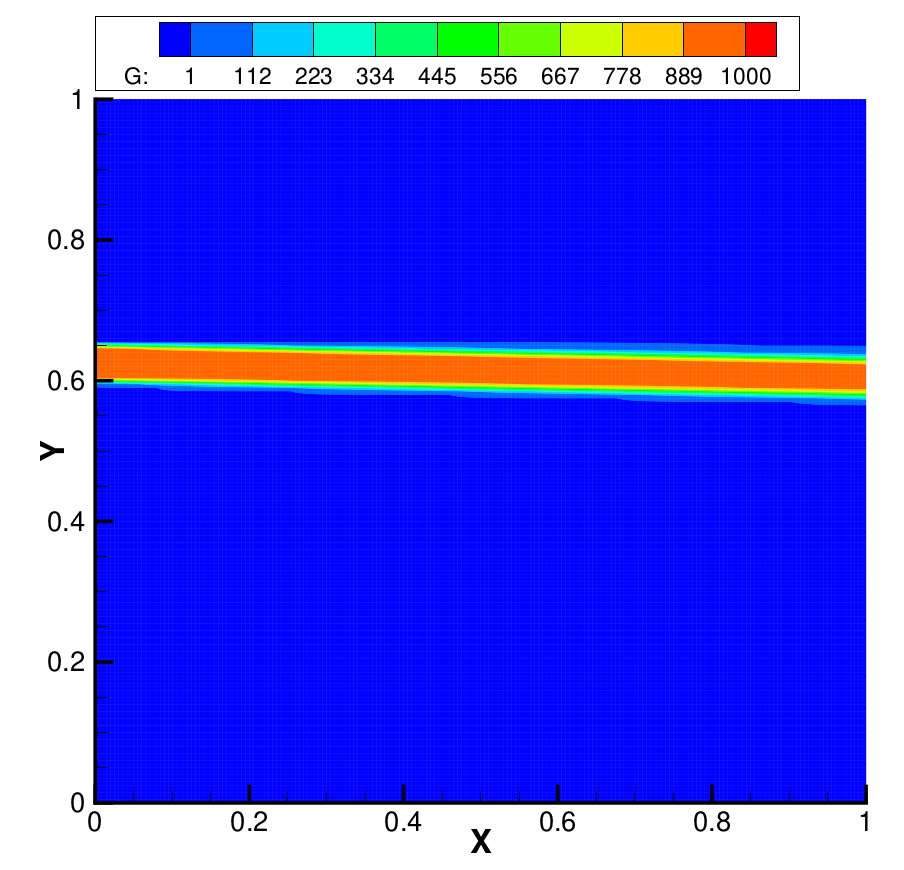}
    \caption{}
    \label{collimated_rayTrav}
  \end{subfigure}\hfill
  \caption{Validation of collimated beam feature (a) geometry (b) contour of irradiation shows the travel of the beam}
\label{collimated_valid_case}
\end{figure}

\begin{figure}[t]
	\centering
	\includegraphics[width=8cm,scale=1]{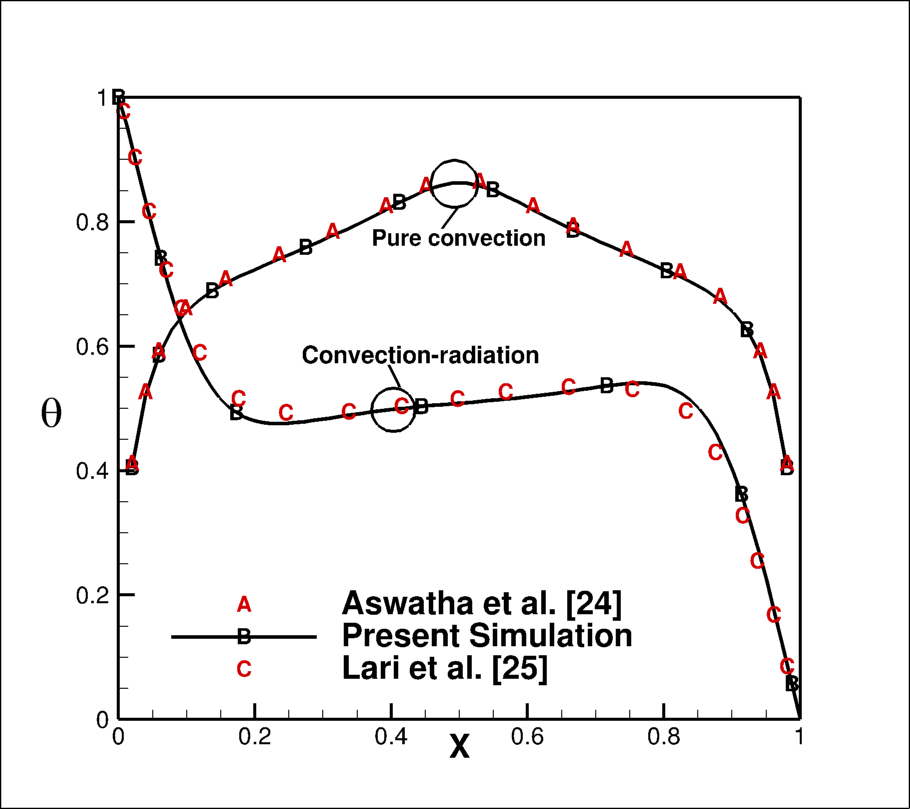}
	\caption{Validation results for pure and combined diffuse radiation with natural convection}
	\label{temp_valid}
\end{figure}

In the absence of any standard benchmark test case for the present problem, the validation has been performed with three cases, first, the standalone feature of collimated beam irradiation problem, in second step, pure natural convection problem which is heated from the bottom, in the third and last step combined convection and radiation in differentially heated cavity have been verified with present solver. The collimated irradiation feature \cite {RAD19} has been tested in a square cavity, as shown in the Fig. (\ref{collimated_geo}). The left side of the wall has a small window of size 0.05 at a height of 0.6. The walls of square cavity are black and cold and also the medium inside the cavity is non-participating. A collimated beam is irradiated on the window in normal direction. It is expected that the beam would travel in normal direction without any attenuation also, the incident should be normal on the opposite wall. Figure (\ref{collimated_rayTrav}) shows the contour of irradiation which clearly shows the travel of collimated in the normal direction without any attenuation. For second step, fluid flow with heat transfer (without radiation) is validated against Aswatha et al. \cite{Aswatha} and combined diffuse radiation and natural convection in a cavity whose top and bottom walls are adiabatic and vertical walls are isothermal at differential temperatures and radiatively opaque, has been validated against Lari et al. \cite{Lari}. Figure \ref{temp_valid} shows the combined results of step two and three and they are in good agreement with the published results.

\section{Grid Independent Test}

Numerical solutions of Navier-Stokes equation and radiation transfer equation is sensitive to the spatial discretization. Additionally, radiative transfer equation also requires angular space discretization, which provides directions where radiation transfer equation is solved. Thus, optimum number of grids and directions have been obtained through independent test study in two steps, as below
\begin{enumerate}
    \item Spatial grids independence test:
    Three spatial grid sizes, i.e 60$\times$60, 80$\times$80 and 100$\times$100 are chosen to calculate the average Nusselt number on the bottom wall for the present problem of natural convection with diffuse radiation. The calculated Nusselt number values for above three grids arrangement on the bottom wall are 6.89, 7.0 and 7.05, respectively. The percentage error between the first and second is 1.52$\%$, whereas between second and third is 0.7$\%$. Thus, the grid points i.e 80$\times$80 is selected for further study.
    \item Angular direction independence test: The polar discretization has no effect on the two-dimensional cases, then OpenFOAM fixed the number of polar directions to 2, in one hemisphere of angular space, while azimuthal considered directions are 3, 5, and 7 for the angular direction independent studied. The total Nusselt numbers on the bottom wall for 80$\times$80 spatial grid and 2$\times$3, 2$\times$5, 2$\times$7 are 6.946, 7.0 and 7.012, respectively. The difference in the second and third angular discreitization is 0.17$\%$. Thus finally $n_\theta \times n_\phi =2 \times 5$ in one hemisphere angular space is selected. 
\end{enumerate}

\section{Results and Discussions}
The present study is performed for natural convection in a cavity that is convective heated from the bottom with heat transfer coefficient 50 $W/m^2$K and free stream temperature 305 K, the side walls are symmetrical cooled at constant temperature 296 K  and top wall is adiabatic. All walls are opaque for radiation. The study is performed in two parts, in first part effect of diffuse radiation on natural convection for various optical thicknesses has been explained. In second part, a semitransparent window of width 0.05 is created on the left wall at height of 0.7 and a collimated beam enters into the cavity through this window at an angle of $45^0$. The effect of collimated beam on fluid flow and heat transfer characteristics for different optical thickness of the fluid has been observed.

\subsection{Combined Diffuse Radiation and Natural Convection}
The effect of various optical thicknesses on fluid flow and heat transfer characteristics have been presented in the section below

\subsubsection{Non-Dimensional Isothermal and Stream Function Contours}

\begin{figure}[t]
\begin{subfigure}{1cm}
    \includegraphics[width=6cm]{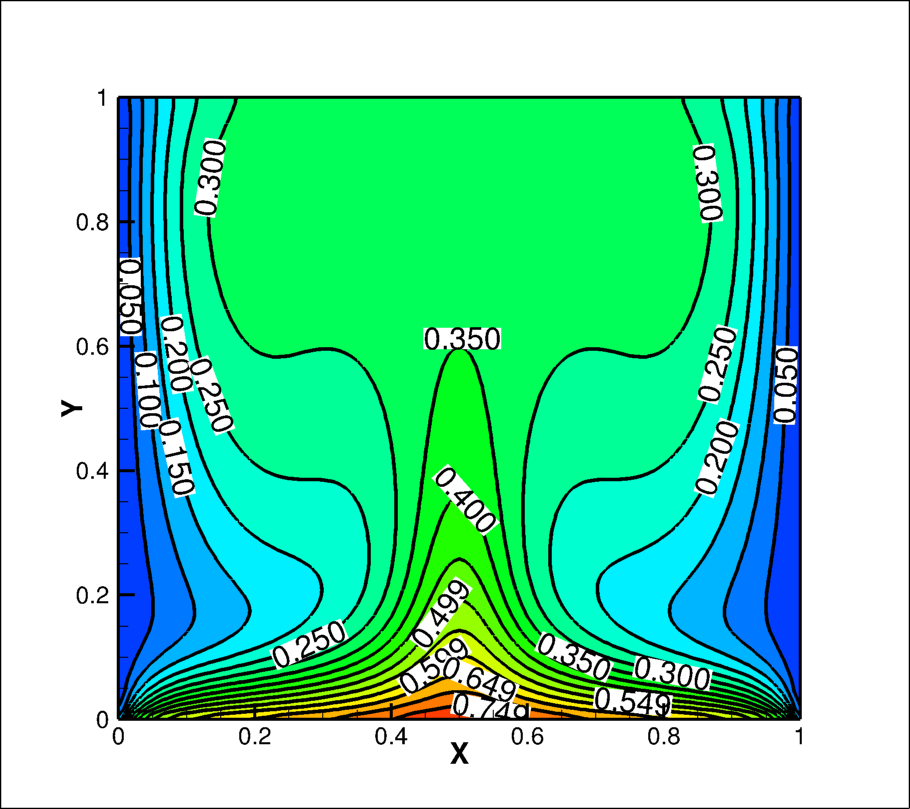}
    \caption{}
    \label{CR_temp}
  \end{subfigure}\hfill
   \begin{subfigure}{6cm}
    \includegraphics[width=6cm]{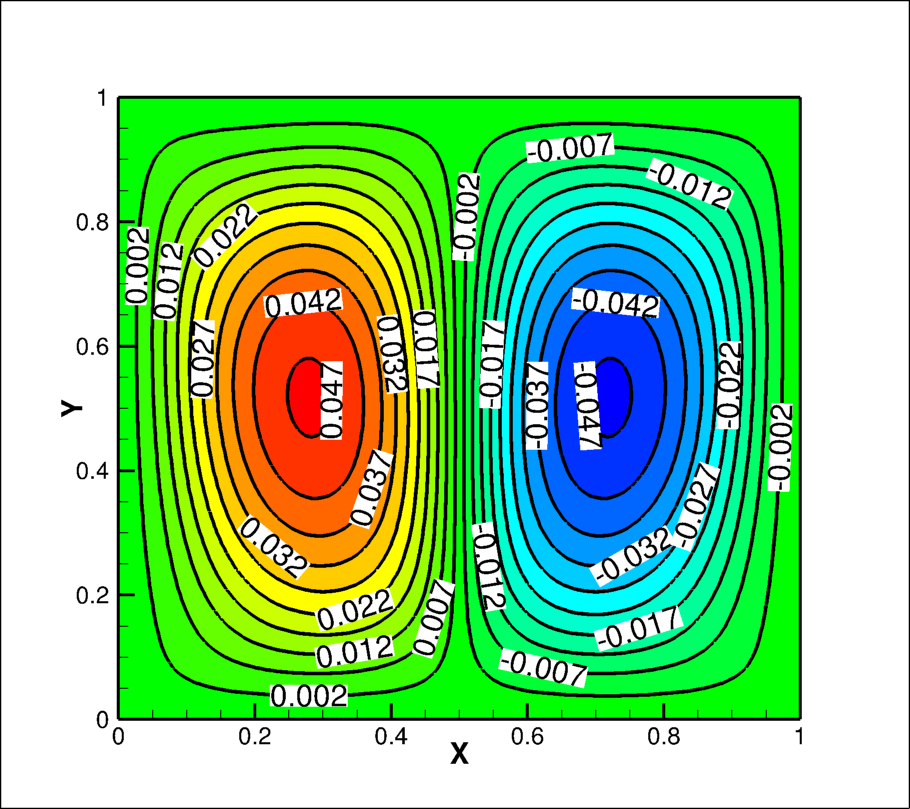}
    \caption{}
    \label{CR_SF}
  \end{subfigure}\hfill
  \caption{Contours of (a)non-dimensional temperature and (b) non-dimensional stream function for optical thickness $\tau=0$ }
\label{diffuse}
\end{figure}

\begin{table}[b]
\centering
\caption{Maximum non-dimensional temperature inside the cavity for various optical thicknesses of the fluid}
\label{ConDif_iso_table}
\begin{tabular}{|ccccccc|}
\hline
\multicolumn{1}{|c}{Optical thickness} & \multicolumn{1}{c}{0} & \multicolumn{1}{c}{0.5} & \multicolumn{1}{c}{1} & \multicolumn{1}{c}{5} & \multicolumn{1}{c}{10} & 50 \\ \hline
Maximum & 0.749 & 0.755 & 0.753 & 0.773 & 0.788 & 0.80 \\ \hline
\end{tabular}
\end{table}
Figure (\ref{CR_temp}) represents the non-dimensional isothermal contours inside the cavity for optical thickness $\tau=0$. The isothermal lines are densely placed near to the bottom wall, sparsely near to the isothermal walls, very less isothermal lines at the central part of the cavity near to the top wall. This reveals the temperature gradients are more near to the bottom wall, less near to isothermal wall and temperature remains uniform at the core of the cavity near to the top wall. The isotherms near to the bottom wall are densely populated, and they originate and rise from both the corners and reach maximum height at the middle of the cavity. This creates a non-uniform temperature and temperature gradient on the bottom wall, thus, non-uniform heating from the bottom. The maximum non-dimensional temperature attains at the mid point of the bottom wall. The non-dimensional isothermal contours are symmetrical about the middle vertical line (i.e $ X=1/2$). This isothermal characteristic is found similar in all optical thicknesses cases (contours are not depicted here), but, the different maximum value of non-dimensional temperatures are shown in Table \ref{ConDif_iso_table}. The maximum non-dimensional temperature increases with the increase of optical thickness of the fluid.

Being a buoyancy driven flow, the temperature field inside the cavity governs the flow field. The contours of non-dimensional stream function of the flow field are depicted in Fig. (\ref{CR_SF}). The bottom wall is non-uniformly heated and temperature field is symmetric about the mid vertical line, this causes to develop two symmetrical counter rotating vortices are established in the flow field in the steady state condition. These two counter rotating vortices are mirror image about the mid-vertical line, unlike the thermal field. The stream line are closely placed at the junction of the two vortices, whereas these lines are relatively sparsely placed in other locations in the cavity, reveals shear-forces are high on the junction of two vortices. The above flow filed characteristics do not change with the change in optical thickness of the fluid. However, the maximum values of non-dimensional stream function change with optical thickness of fluid. Table \ref{ConDif_SF_table} represents the maximum value of non-dimensional stream function with optical thicknesses of the fluid inside the cavity. Also, the maximum non-dimensional stream function values is same for both the vortex, but in opposite sign, representing counter rotating vortices exist in the cavity. The maximum values first increases with optical thicknesses upto 10 then start, decreasing. This reveals long distance phenomenon of radiation changes to local phenomenon with increase of optical thickness of fluid.

\begin{table}[t]
\centering
\caption{Maximum non-dimensional stream function for both the vortices inside the cavity for various optical thicknesses of fluid}
\label{ConDif_SF_table}
\begin{tabular}{|ccccccc|}
\hline
\multicolumn{1}{|c}{Optical thickness} & \multicolumn{1}{c}{0} & \multicolumn{1}{c}{0.5} & \multicolumn{1}{c}{1} & \multicolumn{1}{c}{5} & \multicolumn{1}{c}{10} & 50 \\ \hline
Left vortex & -0.047 & -0.048 & -0.048 & -0.051 & -0.051 & -0.05 \\ \hline
Right vortex & 0.047 & 0.048 & 0.048 & 0.051 & 0.051 & 0.05 \\ \hline
\end{tabular}
\end{table}

\begin{figure}[!htb]
    \centering\includegraphics[width=8cm]{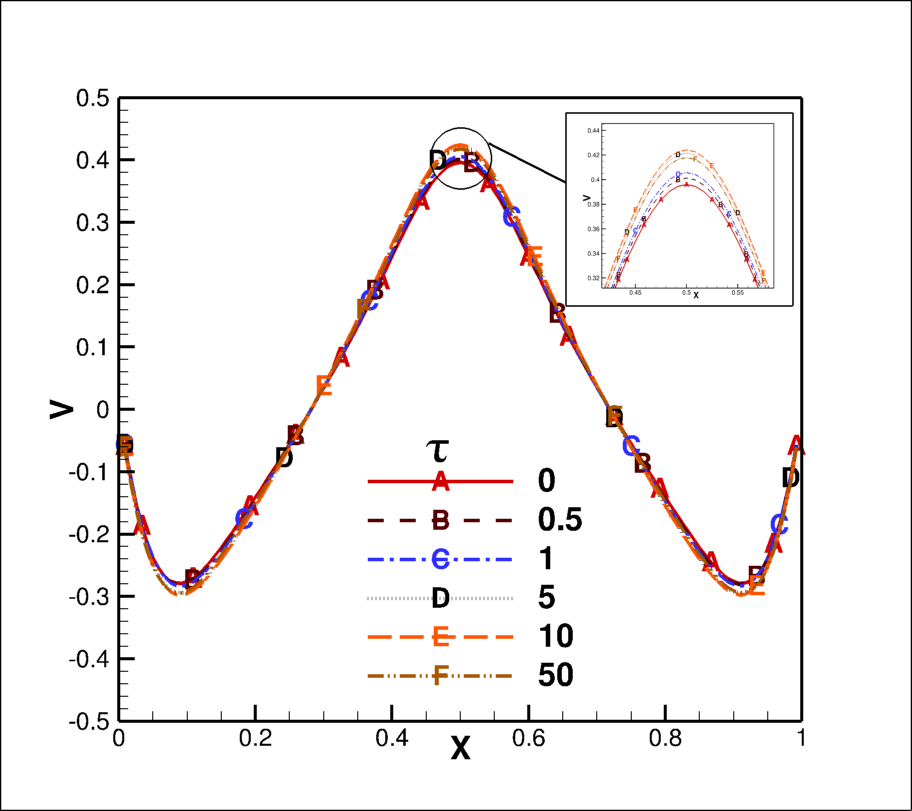}
    \caption{Variation of non-dimensional vertical velocity along the horizontal line at mid height of cavity for various optical thicknesses}
    \label{CR_VV}
 \end{figure}

Figure (\ref{CR_VV}) shows the variation of vertical velocity along the horizontal line at mid height of the cavity. The direction of vertical velocity is in downward direction near to the isothermal walls and reaches a maximum downward velocity at non dimensional distance of 0.1 from both the isothermal walls. Afterwards, the downward velocity starts decreasing and reaches to zero at the centre of both the vortices, i.e, at a distance 0.3 from the isothermal walls, further, the direction of vertical velocity is in upward direction and reaches to peak at the mid point, i.e, the point where two counter-rotating vortices meet. The maximum downward and upward non-dimensional vertical velocities are 0.29 and 0.45, respectively. The non-dimensional vertical velocity curves are same for all optical thicknesses case, however, little variation appears at the inflection point of the curves for different optical thicknesses case as shown in the inset of Fig. (\ref{CR_VV}). The maximum velocity in both the directions increase upto optical thickness 10 and then start decreasing for higher optical thicknesses. 

\begin{figure}[!htb]
 \begin{subfigure}{6cm}
    \includegraphics[width=6cm]{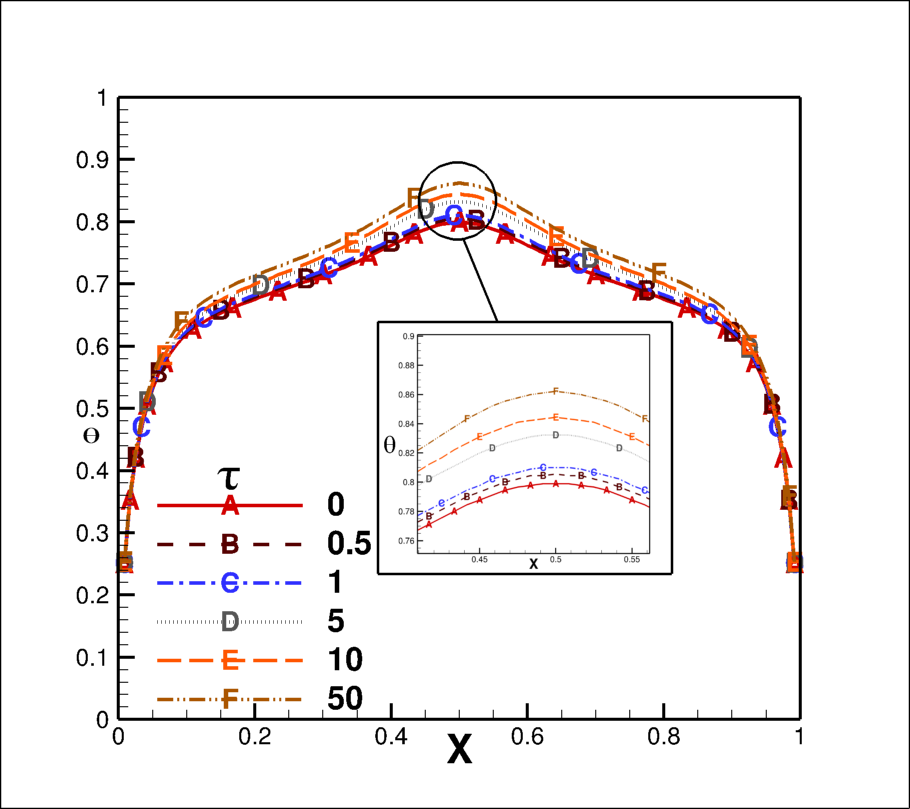}
    \caption{}
    \label{CR_bot_temp}
  \end{subfigure}
   \begin{subfigure}{5cm}
    \includegraphics[width=6cm]{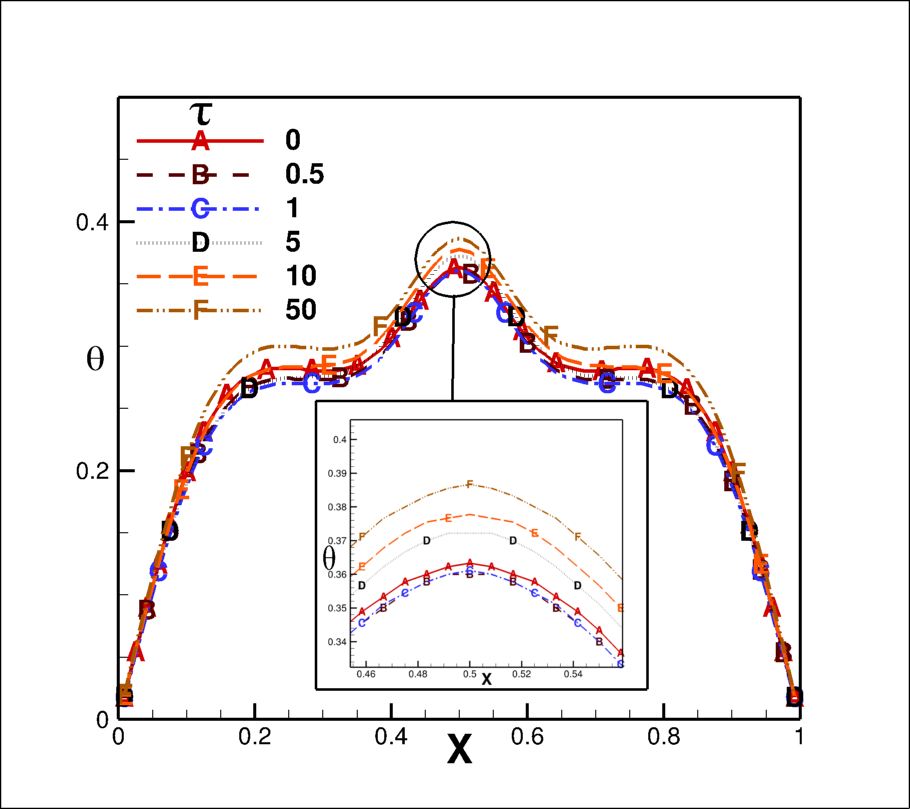}
    \caption{}
    \label{CR_mid_temp}
  \end{subfigure}
   \begin{subfigure}{14cm}
    \centering\includegraphics[width=6cm]{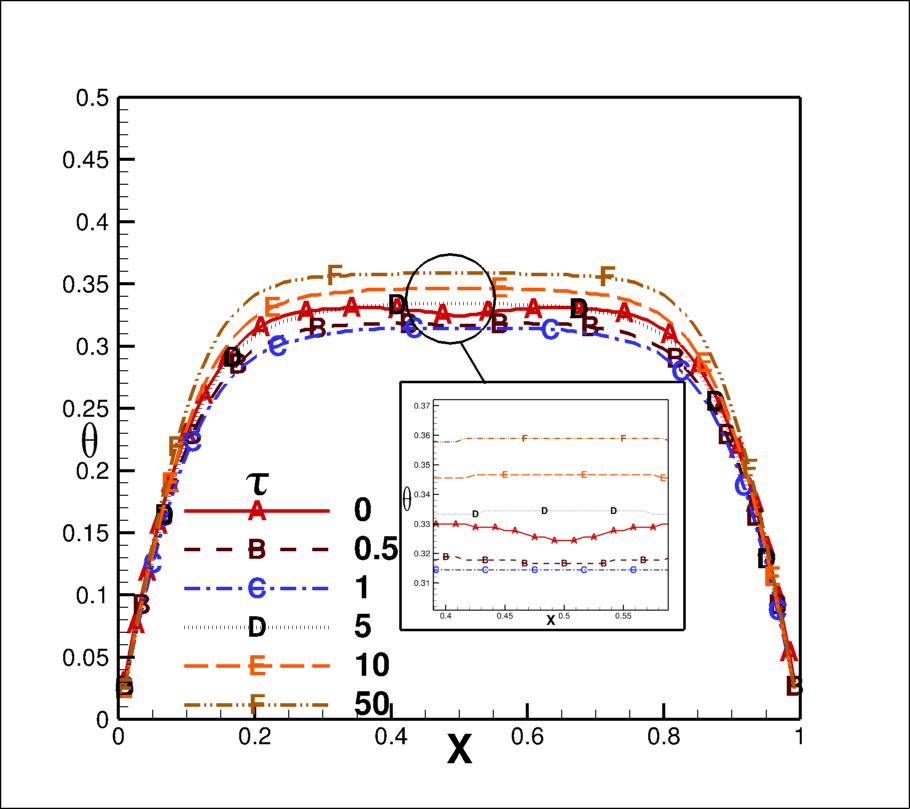}
    \caption{}
    \label{CR_top_temp}
  \end{subfigure}
    \caption{Non-dimensional temperature variation at (a) bottom wall (b) mid- height of cavity and (c) top wall for various optical thicknesses}
\label{CR_wall_temp}
\end{figure}

The non-dimensional temperature variation on the bottom wall along the horizontal direction, at the mid-height of cavity and on the top wall are shown in Fig. \ref{CR_wall_temp} (a), (b) and (c), respectively. The non-dimensional temperature starts sudden increasing from the isothermal walls side upto non-dimensional distance of 0.1 after that the  slow increase in non-dimensional temperature happens and reaches maximum at middle point only on the bottom wall. Whereas, it reaches platue at the non-dimensional distance 0.2 on mid-height of cavity and remain flat on top wall. These curves are again symmetrical about the mid point. The non-dimensional temperature curves remain the same for all optical thicknesses till the first change in the slope reaches, after that temperature curves start showing the difference for different optical thicknesses. The non-dimensional temperature keeps on increasing with optical thicknesses of the fluid.

\subsubsection{Nusselt Number Variation}
\begin{figure}[t]
 \begin{subfigure}{1cm}
    \includegraphics[width=6cm]{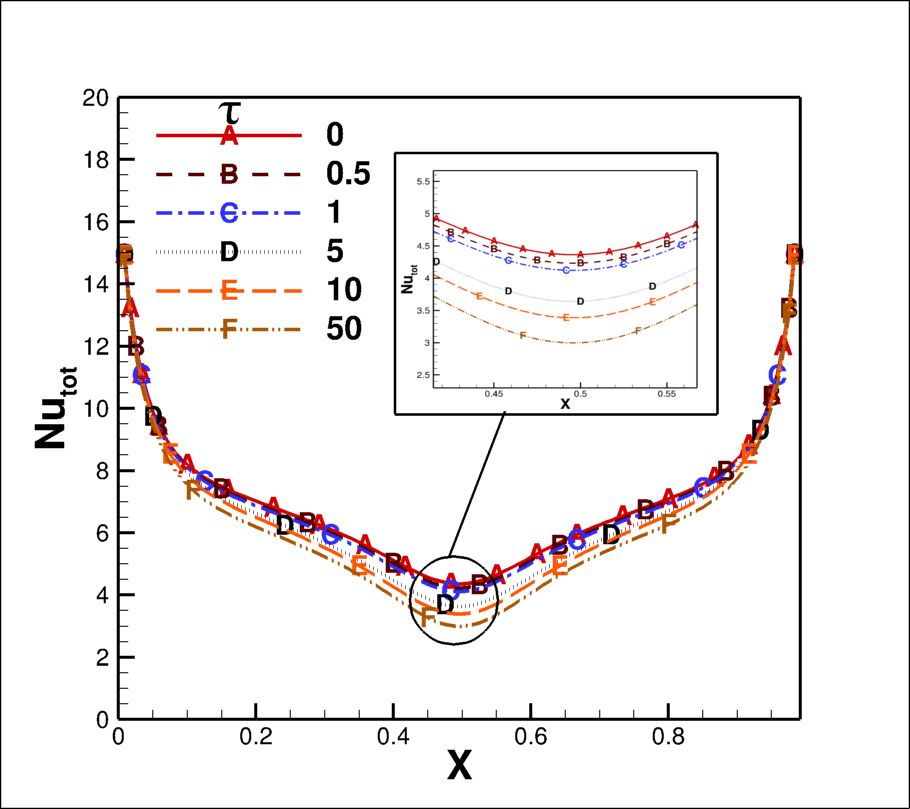}
    \caption{}
    \label{CR_bot_Nu}
  \end{subfigure}\hfill
   \begin{subfigure}{6cm}
    \includegraphics[width=6cm]{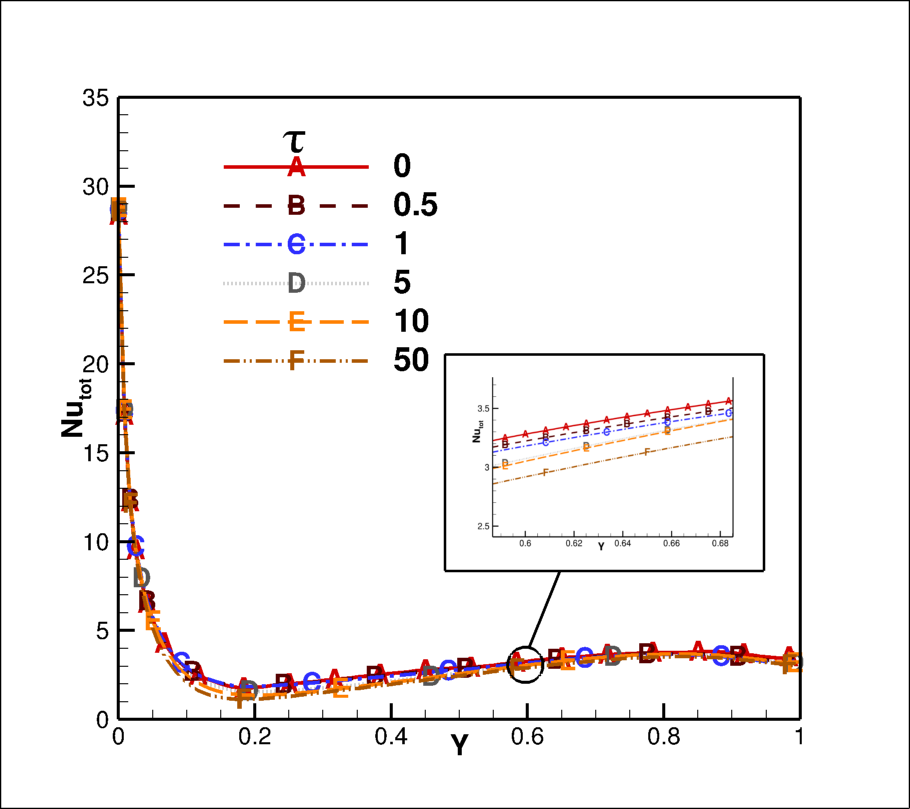}
    \caption{}
    \label{CR_left_Nu}
  \end{subfigure}\hfill
  \caption{Variation of total Nusselt number for different optical thicknesses at (a) bottom and (b) left wall}
\label{CR_Nu}
\end{figure}

The variation of total Nusselt number on the bottom is shown in Fig. (\ref{CR_bot_Nu}). A very high Nusselt number is found near to isotherm walls and this value decreases all off sudden within few distance away from the isothermal walls. Afterwords, the gradient of the curve reduces and finally minimum value of Nusselt number is obtained at mid point of the wall. The lowest value of Nusselt number appears due to the fact that stagnation is at the meeting point of counter rotating vortices, thus a lower temperature gradient. The point of change of gradient of the Nusselt number curve also coincides with the point of maximum value of downward velocity (see Fig. \ref{CR_VV}). The Nusselt number curves are similar for all optical thicknesses near to the isothermal walls till the curves change its gradient. Afterwards, the Nusselt number keeps on decreasing with the increase of optical thicknesses. The maximum value of Nusselt number is 18 near to isothermal walls and minimum is 3 for optical thickness 50 at the midpoint on the bottom wall.

The Nusselt number variation on the isothermal wall is depicted in Fig. \ref{CR_left_Nu}, a very high Nusselt number 28 is found near to bottom of the cavity on the isothermal wall, this Nusselt number decreases very sudden within small height (Y=0.1) from the bottom and rises a little long the height of the wall. The small difference in Nusselt number curve appears at inflection point of curve otherwise, the optical thickness of the fluid has insignificant effect on the Nusselt number on the isothermal walls.

The line average Nusselt number on the bottom and the isothermal walls for the different optical thicknesses are shown in Table \ref{ConDif_avgNu_table}. The conduction Nusselt number is higher than the radiative Nusselt number on both the walls and the conductive Nusselt number increases little with increase in optical thickness whereas for the radiative Nusselt number decreases drastically on the bottom wall this makes total Nusselt number decreases with increase of optical thickness of fluid. A similar observation also been made for the cold wall while Nusselt number is positive on the bottom wall, but, it is negative for the isothermal wall, reveals that the energy enters inside the cavity through bottom and leaves through the isothermal wall.  

\begin{table}[t]
\centering
\caption{Average Nusselt number on different walls for convection diffuse ration case}
\label{ConDif_avgNu_table}
\begin{tabular}{|ccccccl|}
\hline
\multicolumn{1}{|l|}{\multirow{3}{*}{\begin{tabular}[c]{@{}l@{}}Optical \\ thickness ($\tau$)\end{tabular}}} & \multicolumn{6}{c|}{Nusselt number} \\ \cline{2-7} 
\multicolumn{1}{|l|}{} & \multicolumn{3}{c}{Bottom Wall} & \multicolumn{3}{c|}{Cold walls (Left and Right)} \\ \cline{2-7} 
\multicolumn{1}{|l|}{} & \multicolumn{1}{l}{Conduction} & \multicolumn{1}{l}{Radiation} & \multicolumn{1}{l}{Total} & \multicolumn{1}{l}{conduction} & \multicolumn{1}{l}{Radiation} & Total \\ \hline
0 & 5.530 & 1.470 & 7.0 & -2.766 & -0.734 & -3.5 \\ \hline
0.5 & 5.580 & 1.330 & 6.91 & -2.720 & -0.738 & -3.458 \\ \hline
1 & 5.613 & 1.23 & 6.843 & -2.704 & -0.713 & -3.417 \\ \hline
5 & 5.737 & 0.827 & 6.564 & -2.817 & -0.458 & -3.275 \\ \hline
10 & 5.819 & 0.602 & 6.421 & -2.891 & -0.312 & -3.203 \\ \hline
50 & 5.929 & 0.202 & 6.131 & -2.936 & -0.095 & -3.031 \\ \hline
\end{tabular}
\end{table}

\subsection{Combined Collimated/Diffuse Radiation and Natural Convection}
In the previous section, we have seen that the two vortices inside the cavity remain symmetric for all optical thickness of fluid, which may change with collimated irradiation. To simulate collimated irradiation, a semitransparent window of non-dimensional width 0.05 at a non-dimensional height of  0.7 is created on left wall of the cavity and a collimated beam irradiated at an angle of $45^0$. The beam travels into the fluid in that may have different optical thicknesses and the dynamics of these two vortices and heat transfer characteristics have been studied with collimated irradiation.
  
\subsubsection{Irradiation Contours}

\begin{figure}[!hbt]
\begin{subfigure}{4cm}
    \includegraphics[width=6cm]{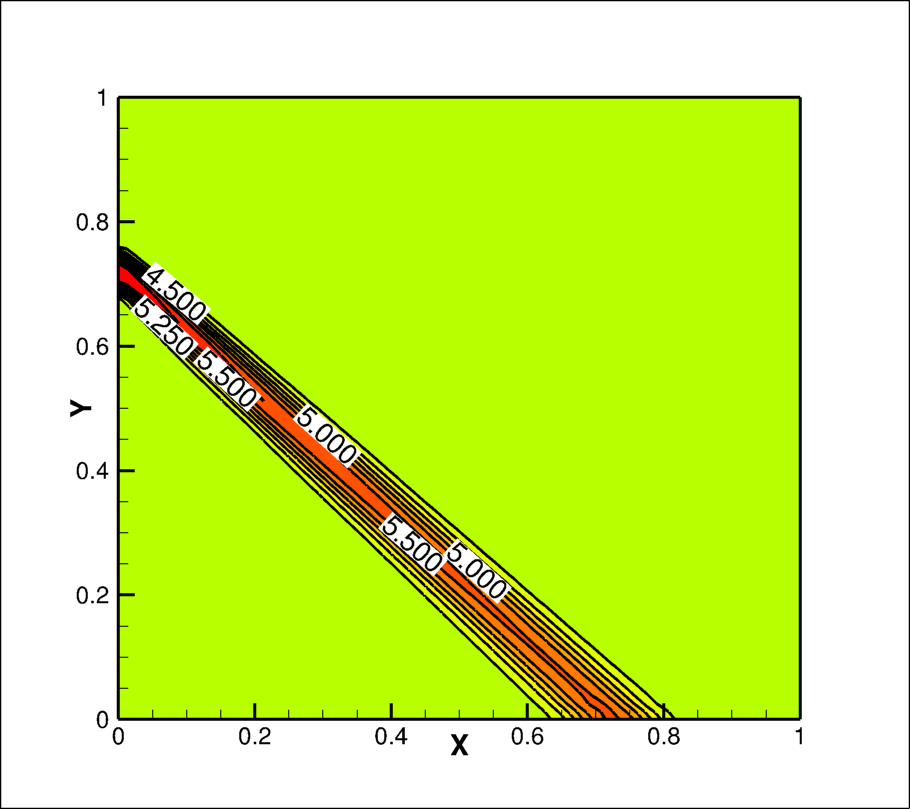}
    \caption{}
    \label{G_C_N0}
  \end{subfigure}
   \begin{subfigure}{5cm}
    \includegraphics[width=6cm]{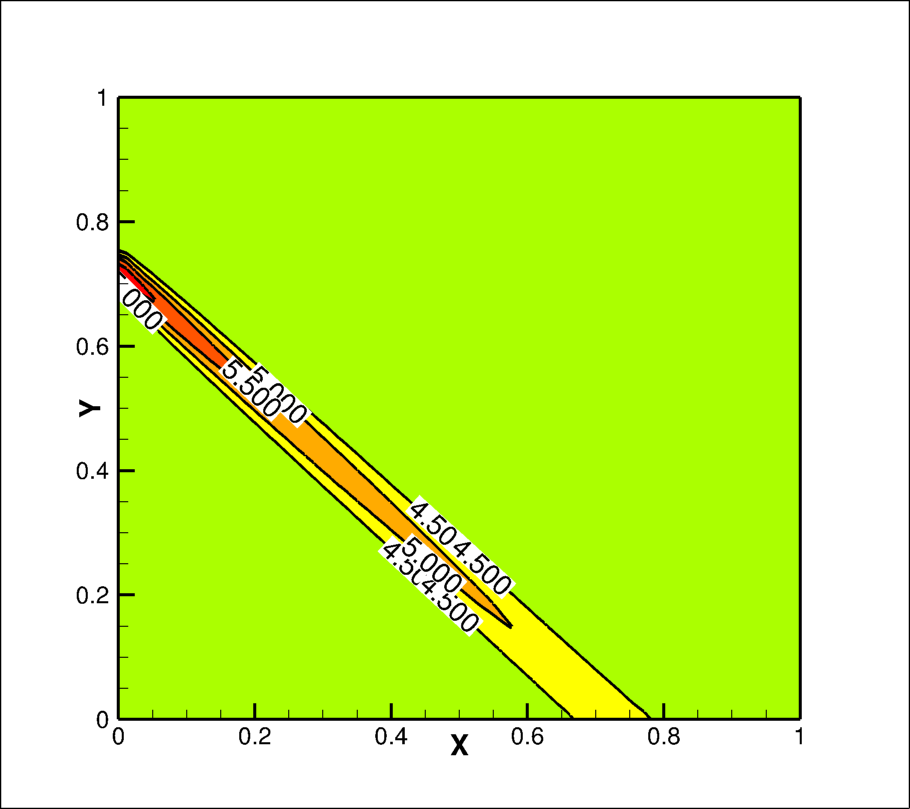}
    \caption{}
    \label{G_C_NP5}
  \end{subfigure}
  \begin{subfigure}{5cm}
    \includegraphics[width=6cm]{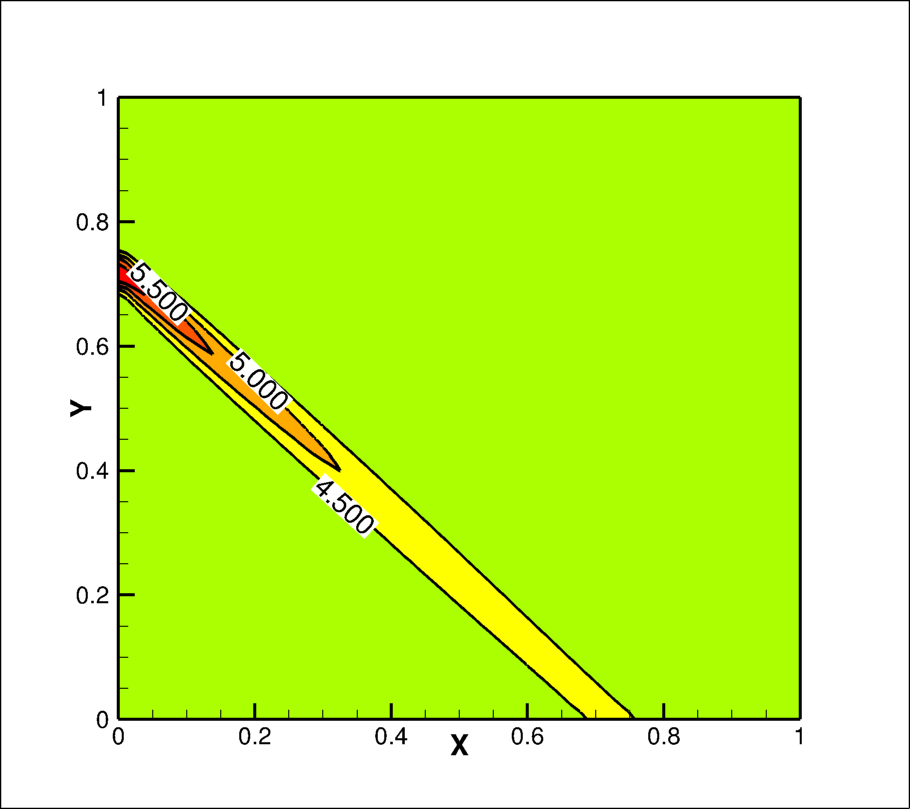}
    \caption{}
    \label{G_C_N1}
  \end{subfigure}
   \begin{subfigure}{5cm}
    \includegraphics[width=6cm]{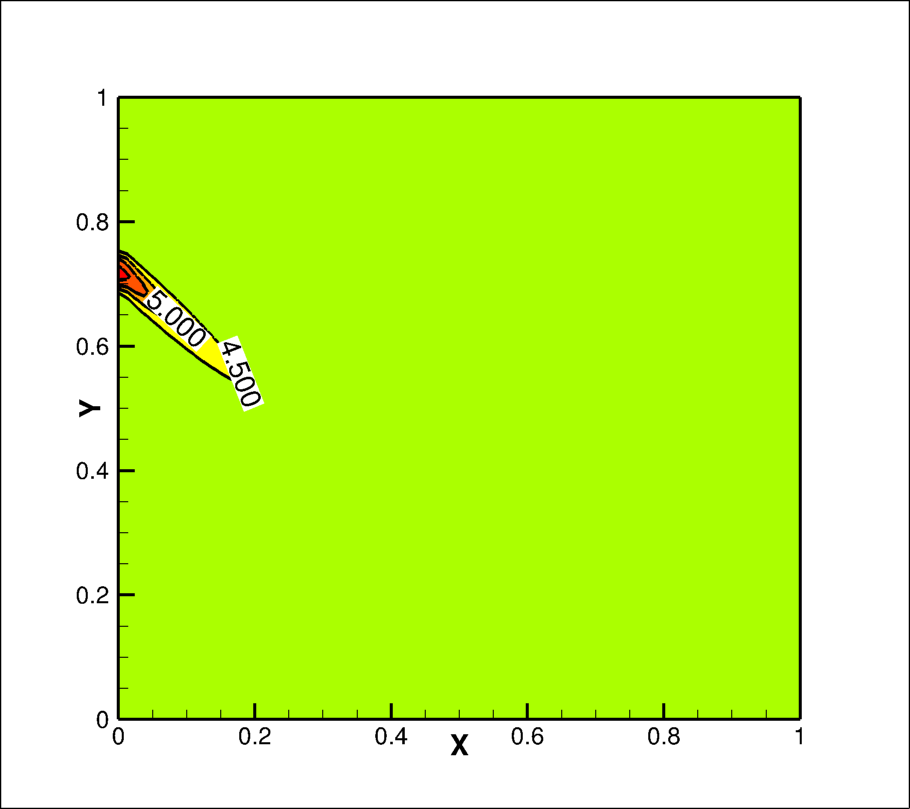}
    \caption{}
    \label{G_C_N5}
  \end{subfigure}
  \begin{subfigure}{5cm}
    \includegraphics[width=6cm]{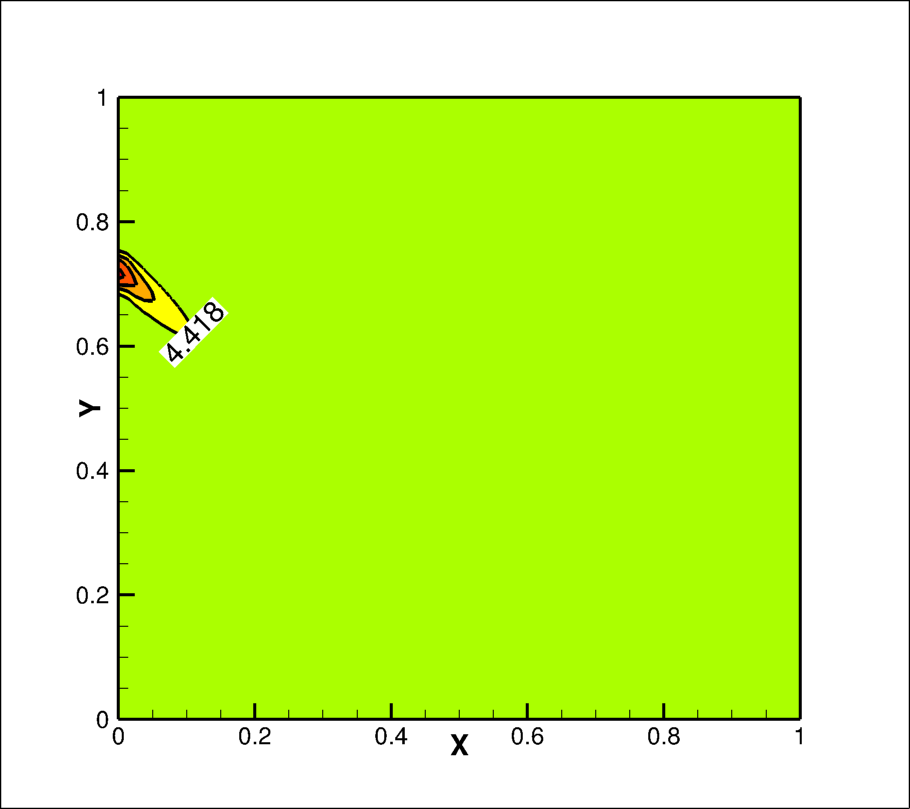}
    \caption{}
    \label{G_C_N10}
  \end{subfigure}
  \hspace{2.0cm}
  \begin{subfigure}{5cm}
    \includegraphics[width=6cm]{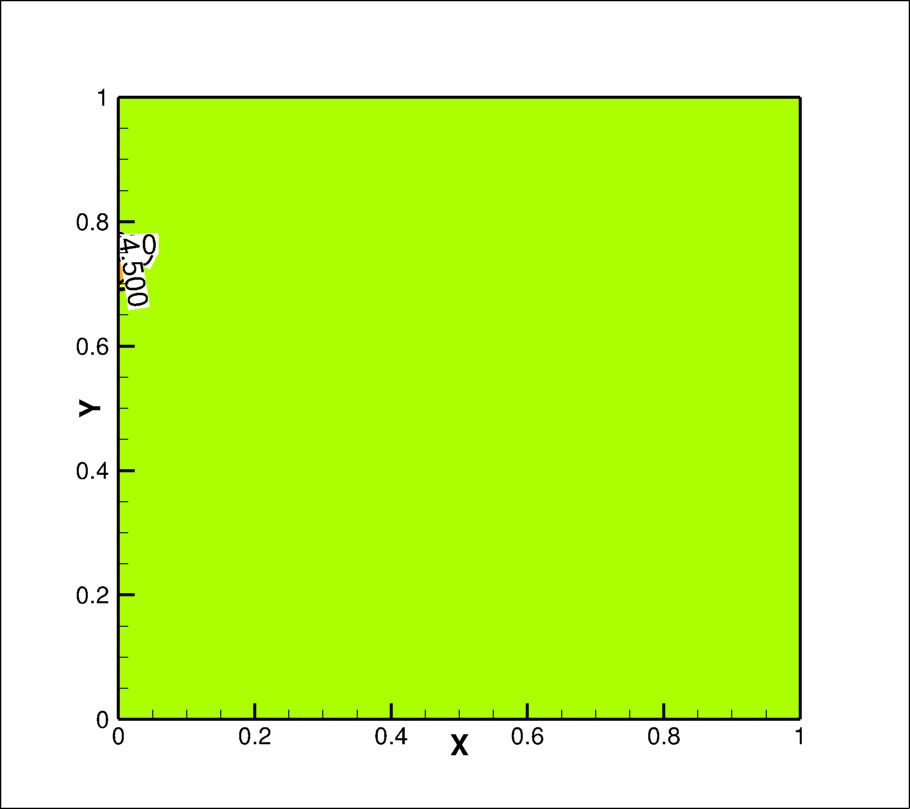}
    \caption{}
    \label{G_C_N50}
  \end{subfigure}
  \caption{Contours of Non-dimensional irradiation for various optical thicknesses of (a) $\tau=0$ (b) $\tau=0.5$ (c) $\tau=1$ (d) $\tau=5$ (e) $\tau=10$ and (f)$\tau=50$}
\label{G_collimated}
\end{figure}

Progression of the collimated beam inside the cavity can be best represented by irradiation contours. The non-dimensional irradiation contours inside the cavity for optical thickness ($\tau$) 0, 0.5, 1, 5, 10 and 50 of the fluid are shown in Fig.  \ref{G_collimated}(a)-(f), respectively. The optical thickness $\tau=0$ corresponds to transparent medium i.e no absorption-emission by fluid, therefore, the irradiation strength remains constant till it reaches to the bottom wall, here at non-dimensional  distance of 0.7 from the left corner on bottom wall, whereas, strength of collimated beam reduces for non-zero optical thickness of the fluid. This decrement increases with optical thickness as shown in Fig. \ref{G_collimated}(b)-(f). The collimated irradiation does not reach to the bottom wall for optical thickness 5 of fluid onwards. Furthermore the collimated energy gets absorbed near to the window for optical thickness 50 (see Fig. \ref{G_collimated}f).

\subsubsection{Stream lines and Isothermal line contours}

\begin{figure}[!htb]
 \begin{subfigure}{4cm}
    \centering\includegraphics[width=6cm]{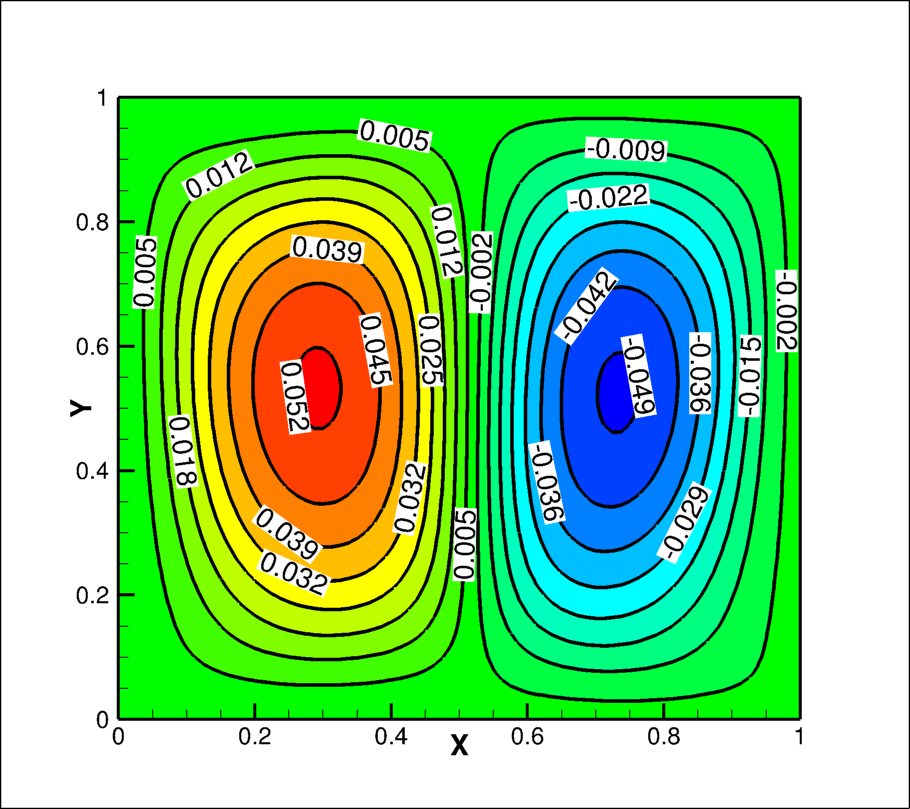}
    \caption{}
    \label{SF_C_N0}
  \end{subfigure}
   \begin{subfigure}{5cm}
    \centering\includegraphics[width=6cm]{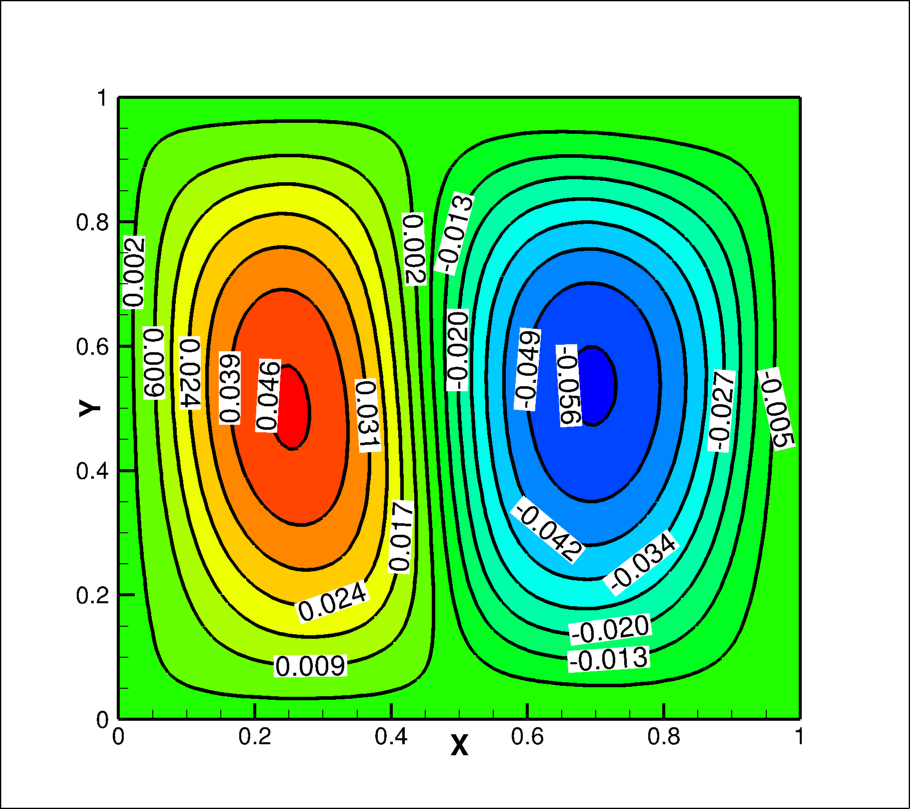}
    \caption{}
    \label{SF_C_NP5}
  \end{subfigure}
  \begin{subfigure}{5cm}
    \centering\includegraphics[width=6cm]{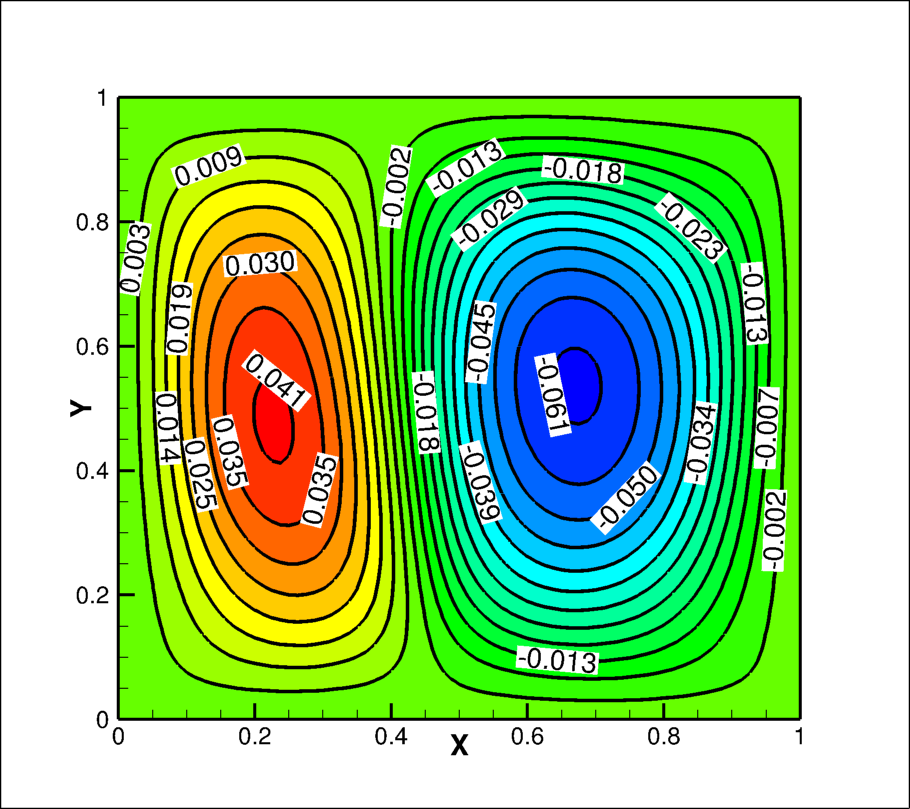}
    \caption{}
    \label{SF_C_N1}
  \end{subfigure}
   \begin{subfigure}{5cm}
    \centering\includegraphics[width=6cm]{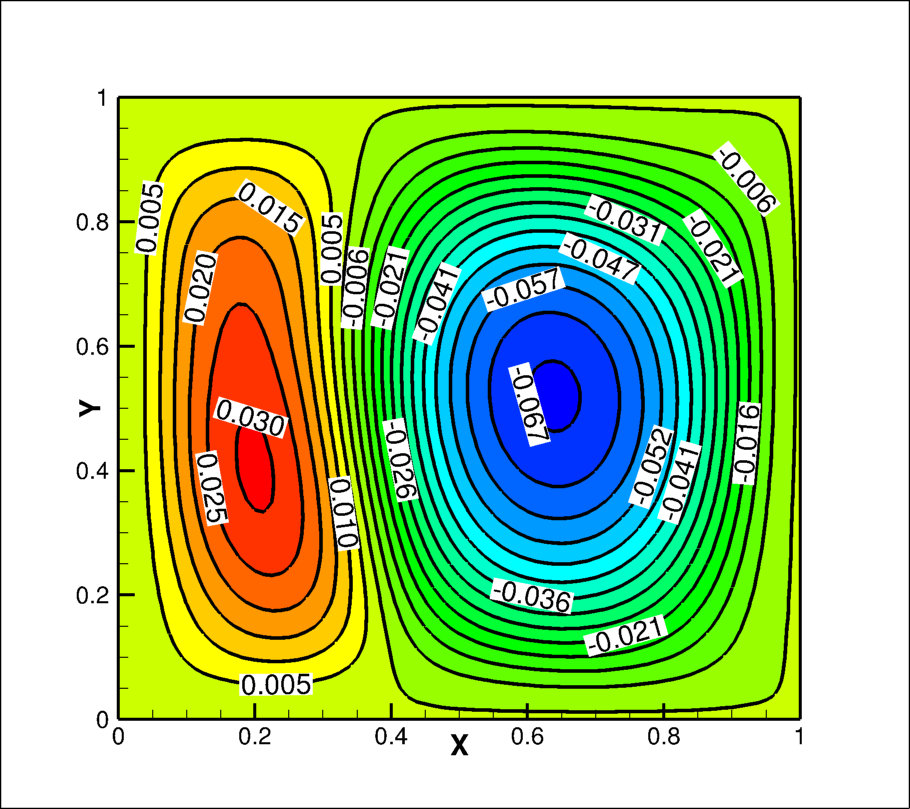}
    \caption{}
    \label{SF_C_N5}
  \end{subfigure}
  \begin{subfigure}{5cm}
    \centering\includegraphics[width=6cm]{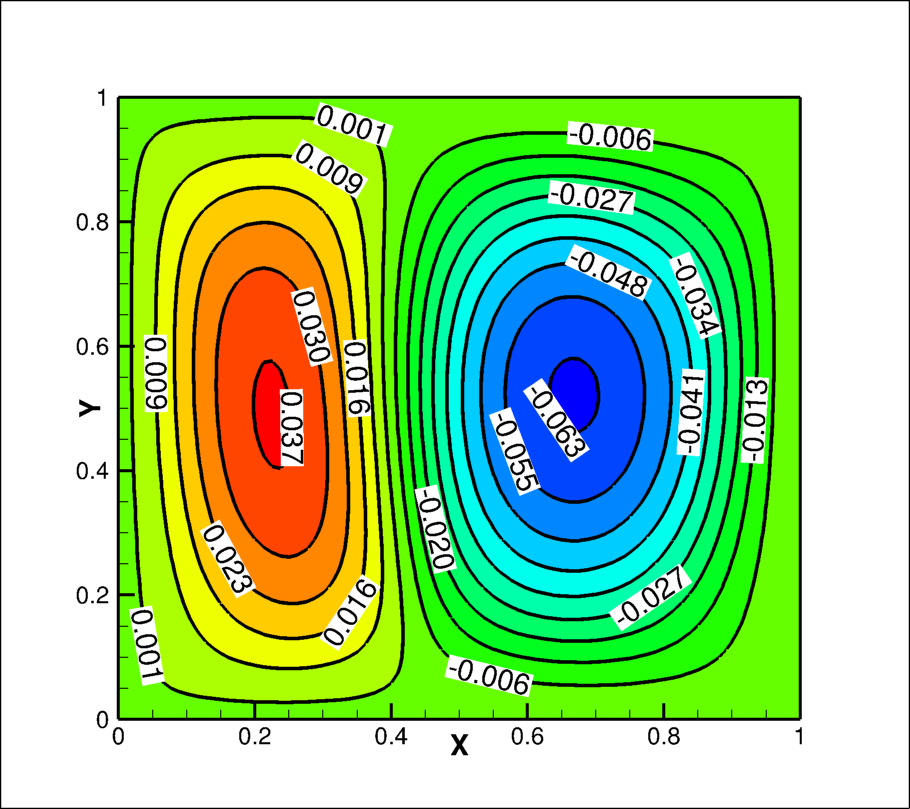}
    \caption{}
    \label{SF_C_N10}
  \end{subfigure}
  \hspace{2.0cm}
  \begin{subfigure}{5cm}
    \centering\includegraphics[width=6cm]{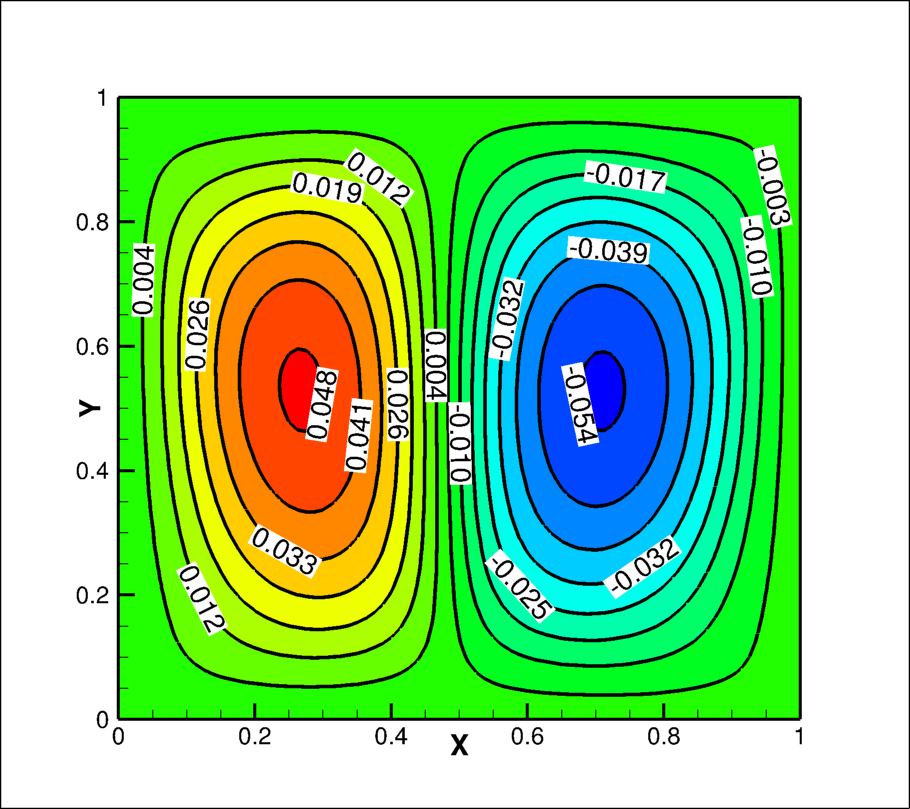}
    \caption{}
    \label{SF_C_N50}
  \end{subfigure}
  \caption{Contours of Non-dimensional stream function for various optical thicknesses of (a) $\tau=0$ (b) $\tau=0.5$ (c) $\tau=1$ (d) $\tau=5$ (e) $\tau=10$ and (f)$\tau=50$}
\label{SF_collimated}
\end{figure}

The non-dimensional stream function contours inside the cavity for optical thickness 0, 0.5, 1, 5 10 and 50 are shown in Fig. \ref{SF_collimated}(a)-(f), respectively. As we have seen in previous section, the cavity contains two symmetrical right and left vortices occupying equally half of the space inside cavity without collimated incidence, the collimated incidence causes asymmetricity in these two vortices which vary with behaviour of the fluid for the radiation heat transfer. As the fluid is transparent for radiation transfer, all the collimated beam energy strikes on the bottom wall at a non-dimensional distance of 0.7 from left corner and creates a hot spot on the wall. There is enhance in buoyancy force in upward direction at that location, this inturn, enhances the upwards direction force in the right vortex, therefore the right vortex becomes thinner and left vortex becomes thicker. The opposite behaviour has been observed with non-zero optical thickness in the fluid. This may be owing to fact that collimated beam is travelling through the left vortex which absorbs the radiation energy and creates local heating of fluid, this enhances upward buoyancy force in the left vortex, whereas some heat is also  being transfered to right vortex through absorption in the right vortex and hot spot at the bottom. The energy absorbed by left vortex may be higher due to larger distance travelled by collimated beam in left vortex, this causes decrease in the size of left vortex and increases in size of right vortex.

The above scenario for optical thickness continues upto $\tau=10$ of the fluid, however, this changes with further increase of the optical thickness of the fluid. Further,due to the increase in the optical thickness, the size of left and right vortices keeps on increasing and decreasing, respectively. This can be infer from the irradiation contours (Fig. \ref{G_collimated}e and f) that all collimated energy gets absorbed near to the wall only and most of the energy is transferred out of domain due to isothermal condition of the wall, instead of transferring to the fluid. Thus, there is relative decrease in the upward buoyancy force and result in little increase in size of left vortex. There is further increase in the size of left vortex with increase in optical thickness of the fluid. Nevertheless, a situation may arrive with very very high optical thickness case when collimated irradiation case  will be similar to without collimated beam case.

\begin{table}
\centering
\caption{Maximum non-dimensional stream function value for various optical thicknesses of fluid}
\label{co_SF_table}
\begin{tabular}{|ccccccc|}
\hline
Optical thickness & 0 & 0.5 & 1 & 5 & 10 & 50 \\ \hline
\begin{tabular}[c]{@{}c@{}} Left vortex \end{tabular} & 0.052 & 0.046 & 0.041 & 0.03 & 0.04 & 0.048 \\ \hline
Right vortex & -0.049 & -0.056 & -0.061 & -0.07 & -0.06 & -0.054 \\ \hline
\end{tabular}
\end{table}

Table \ref{co_SF_table} shows the value of maximum non-dimensional stream function for left and right vortices with optical thickness of the fluid. The non-dimensional maximum value of stream function of left vortex decreases with the increases of optical thickness of fluid upto 5, then increases. Whereas, this value for right vortex increases upto the optical thickness 5, then decreases. The negative stream function value indicates that direction of rotation of vortex is in clock wise.

\begin{figure}[!htb]
 \begin{subfigure}{4cm}
    \centering\includegraphics[width=6cm]{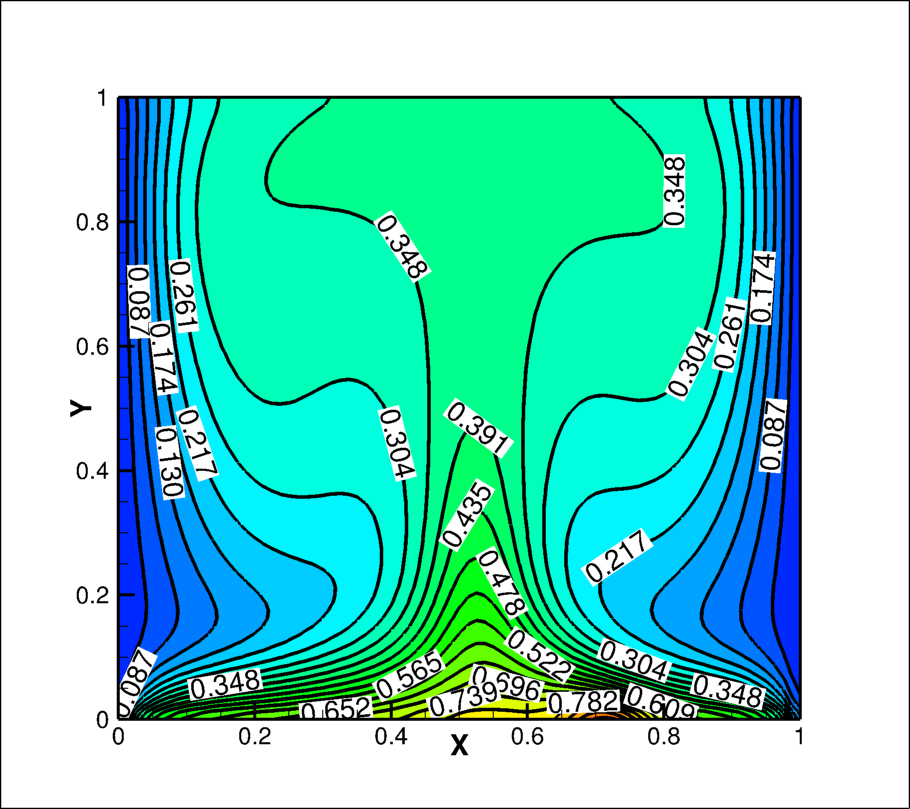}
    \caption{}
    \label{Temp_C_N0}
  \end{subfigure}
   \begin{subfigure}{5cm}
    \centering\includegraphics[width=6cm]{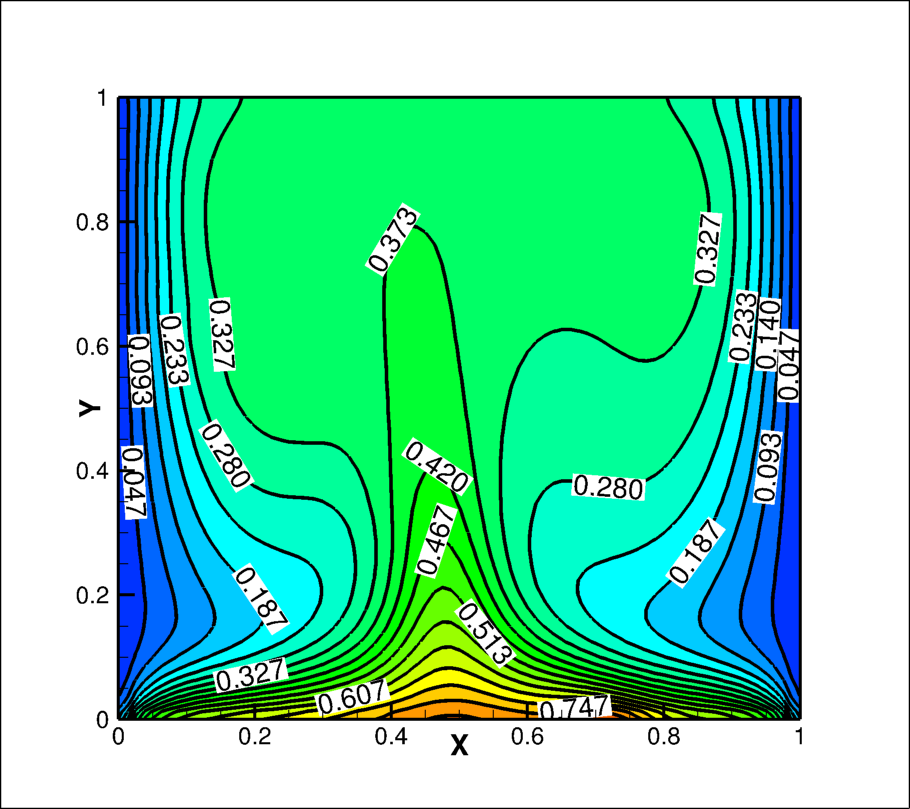}
    \caption{}
    \label{Temp_C_NP5}
  \end{subfigure}
  \begin{subfigure}{5cm}
    \centering\includegraphics[width=6cm]{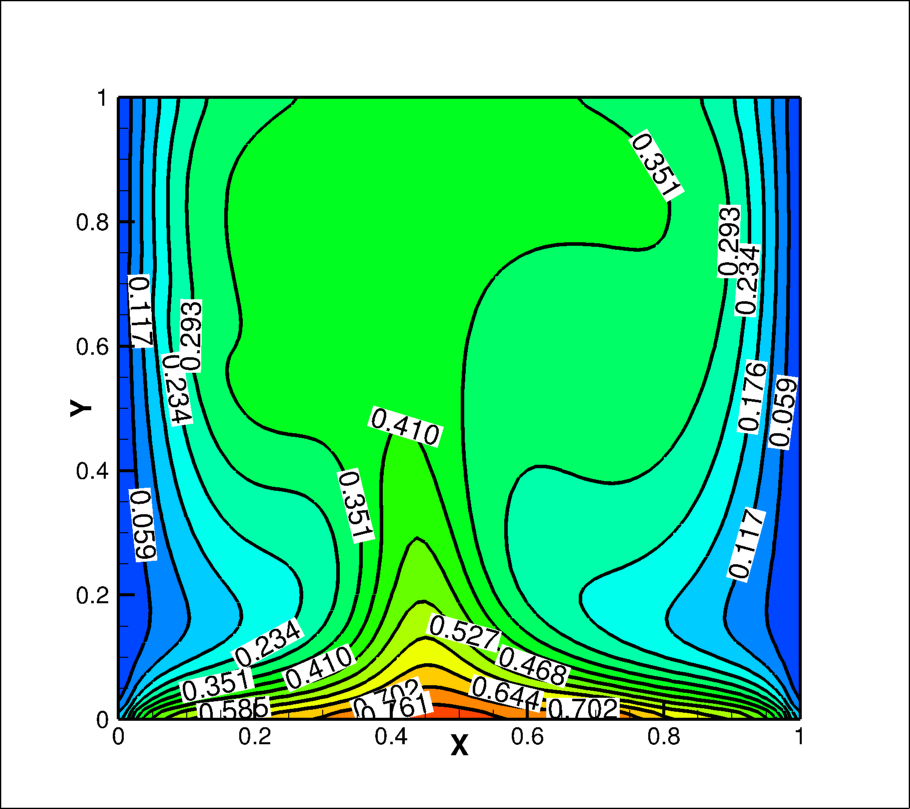}
    \caption{}
    \label{Temp_C_N1}
  \end{subfigure}
   \begin{subfigure}{5cm}
    \centering\includegraphics[width=6cm]{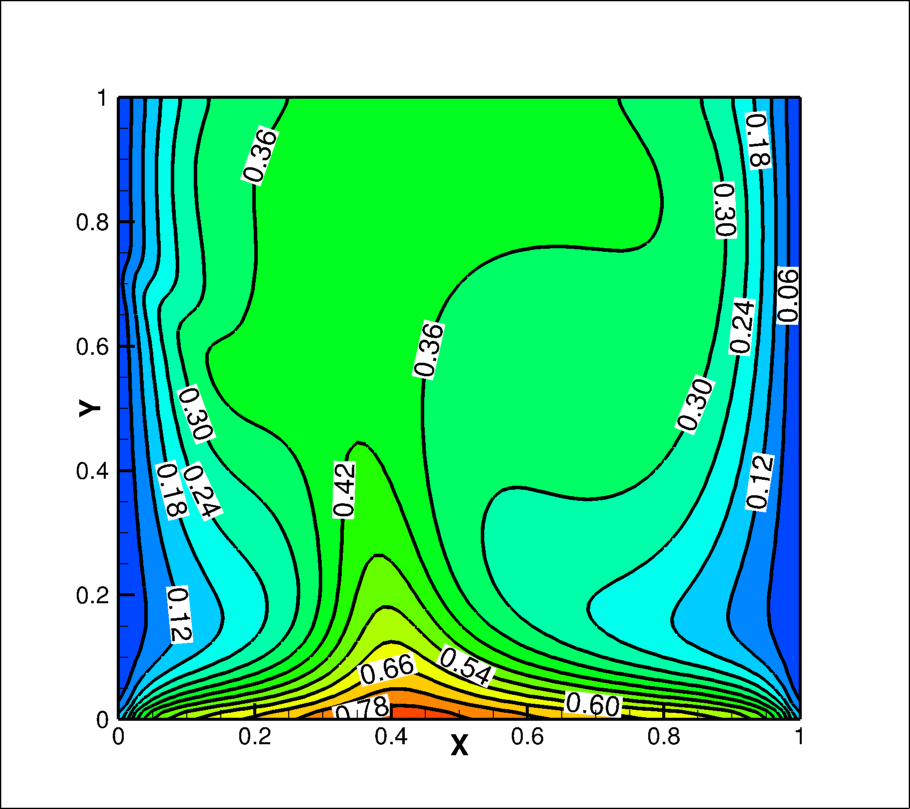}
    \caption{}
    \label{Temp_C_N5}
  \end{subfigure}
  \begin{subfigure}{5cm}
    \centering\includegraphics[width=6cm]{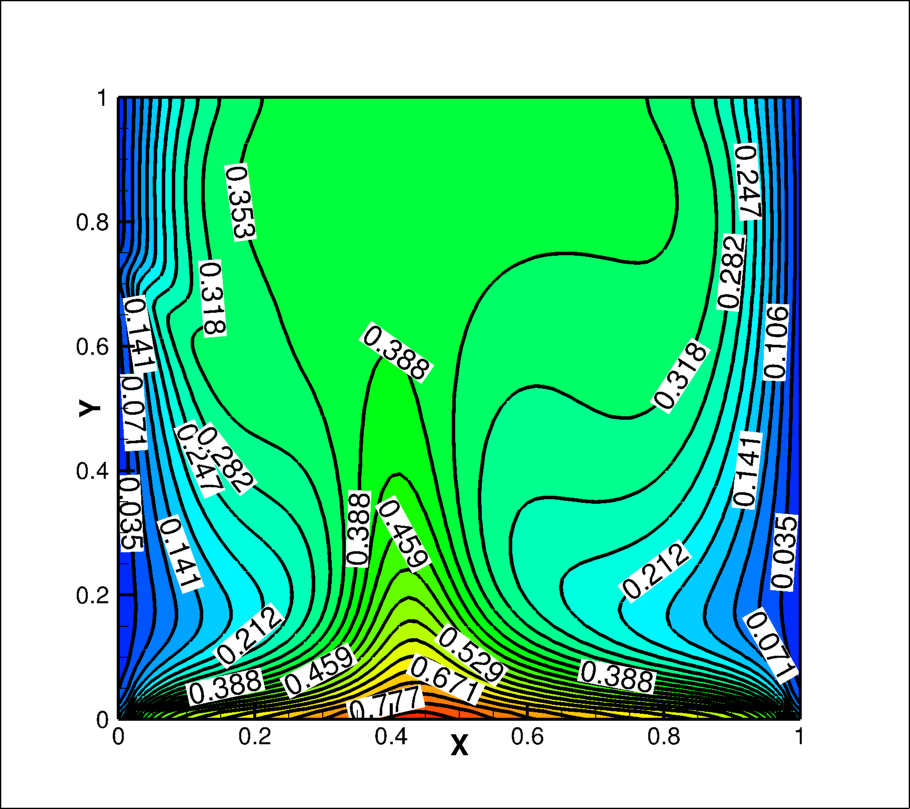}
    \caption{}
    \label{Temp_C_N10}
  \end{subfigure}
  \hspace{2.0cm}
  \begin{subfigure}{5cm}
    \centering\includegraphics[width=6cm]{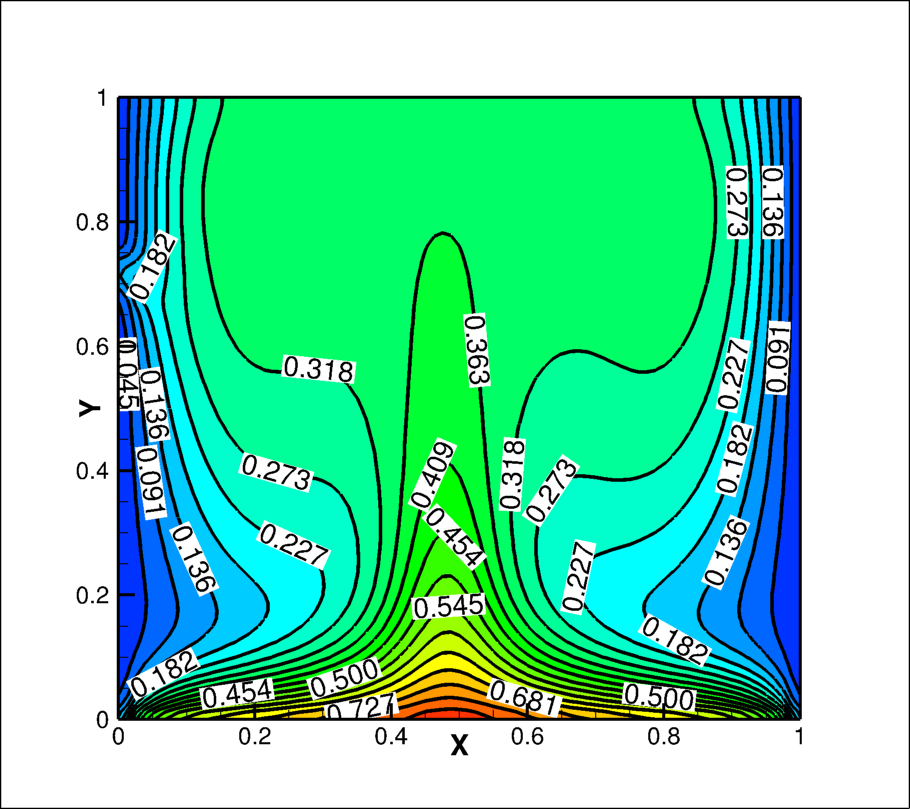}
    \caption{}
    \label{Temp_C_N50}
  \end{subfigure}
  \caption{Contours of non-dimensional isothermal for various optical thicknesses of (a) $\tau=0$ (b) $\tau=0.5$ (c) $\tau=1$ (d) $\tau=5$ (e) $\tau=10$ and (f)$\tau=50$}
\label{Temp_collimated}
\end{figure} 

The effect of collimated beam irradiation on the isothermal contours inside the cavity for optical thickness 0, 0.5, 1, 5, 10 and 50  are shown in Fig. \ref{Temp_collimated}(a)-(f), respectively. In absence of collimated irradiance the symmetrical isotherm about the vertical line at middle of the cavity  gets tilted either right or left to the vertical line depending upon the behaviour of the fluid for radiation energy. The isothermal lines are tilted towards right for transparent fluid, (see Fig. \ref{Temp_collimated}(a)), this is due to the fact that right vortex is smaller than the left vortex and reverse is true for non-zero optical thickness fluid (see Fig. \ref{Temp_collimated}(b)-(c)) i.e, the isotherms are tilted towards left. The area near to top adiabatic wall away from isothermal walls is almost at uniform temperature at lower optical thickness, however, the area of uniform temperature reduces with higher optical thickness and also, the isotherm lines do not remain smooth. The isotherm lines are bent in the line of collimated beam at higher optical thickness of fluid. This bend of isothermal lines diminishes with increase of optical thicknesses and remain limited near to the wall for optical thickness 50. Table \ref{co_iso_table} represents the maximum non-dimensional temperature that exists in side cavity. The maximum non-dimensional temperature decreases with the increase of optical thickness of the fluid upto 1 then starts increasing and this maximum non-dimensional temperatures always exist on the bottom wall (see Fig. \ref{Temp_collimated}).
\begin{figure}[t]
    \centering\includegraphics[width=8cm]{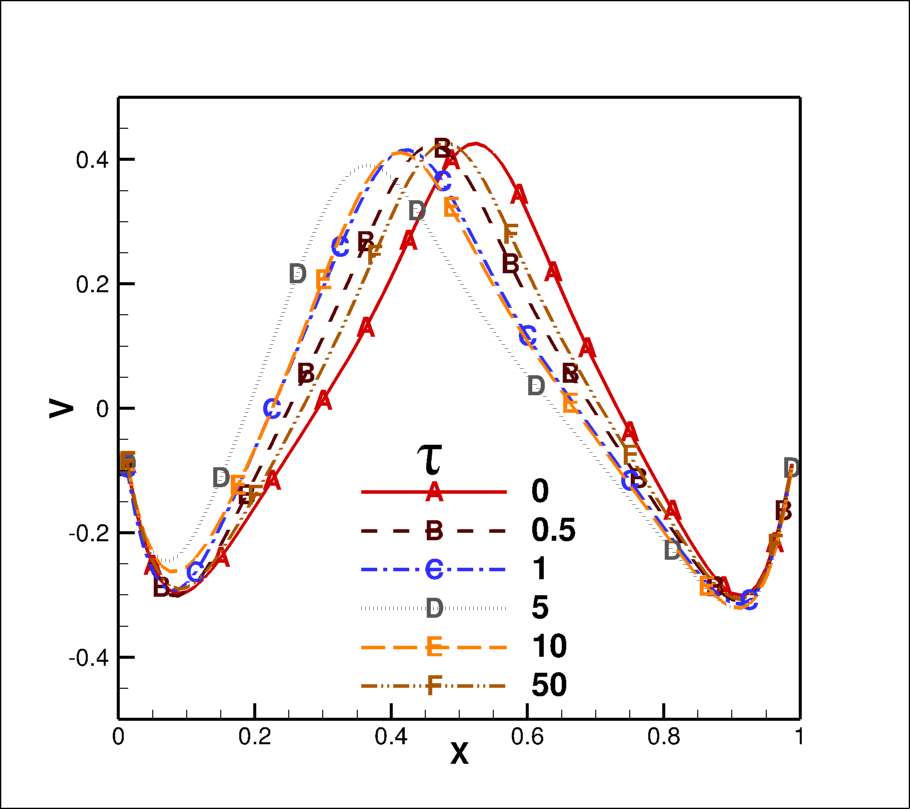}
     \caption{Variation of non-dimensional vertical velocity along the horizontal line at the mid height of the cavity}
    \label{C_CR_VV}
\end{figure} 

\begin{table}
\centering
\caption{Maximum non-dimensional temperature inside the cavity for various optical thicknesses of fluid}
\label{co_iso_table}
\begin{tabular}{|ccccccc|}
\hline
Optical thickness & 0 & 0.5 & 1 & 5 & 10 & 50 \\ \hline
Maximum & 0.894 & 0.760 & 0.761 & 0.78 & 0.78 & 0.801 \\ \hline
\end{tabular}
\end{table}

The variation of non-dimensional vertical velocity in the horizontal direction at the mid height of the cavity is shown in Fig. \ref{C_CR_VV}. The vertical velocity is non-dimensionalised by convective velocity scale as explained in section 3.1. The vertical velocity starts increasing in downward direction and reaches to maximum value at 0.065 distance from wall and 0.09 distance away from the right wall for all optical thickness of the fluid, then its value starts decreasing and reaches to zero at centre point of both the vortices, afterwards, its direction is in upward and again reaches to maximum value. The non-dimensional maximum upward velocity is achieved at different locations for different optical thicknesses of medium, and this location is at junction of the two vortices. It is therefore, the location of maximum upward velocity is little bit towards right to mid point for transparent fluid and left for non-zero optical thickness of fluid, whereas, the location of maximum downward velocity remain almost same for all optical thicknesses. The value of maximum upward velocity also decreases for optical thickness of fluid greater than 5. This could be the owing to fact that less energy is being transferred to the fluid from the bottom wall for case of optical thickness greater than 1. The maximum downward non-dimensional velocity is 0.3 and upwards velocity is 0.4 for optical thickness 0 and minor difference in these values for other optical thicknesses. 

\begin{figure}[t]
 \begin{subfigure}{6cm}
    \includegraphics[width=6cm]{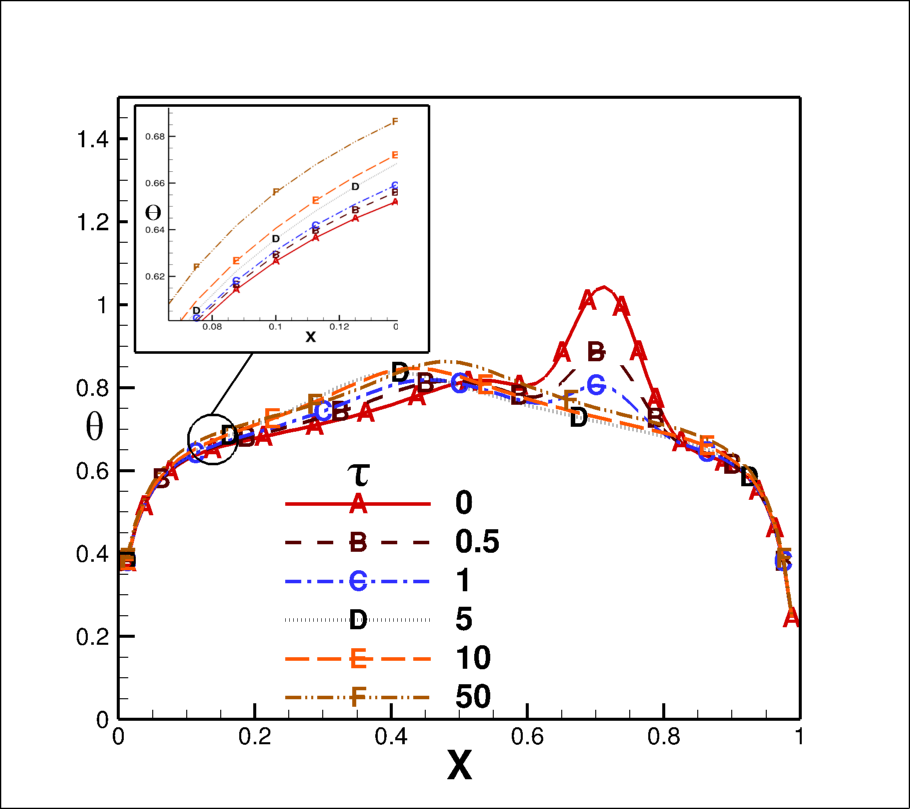}
    \caption{}
    \label{C_CR_BT}
  \end{subfigure}
   \begin{subfigure}{5cm}
   \includegraphics[width=6cm]{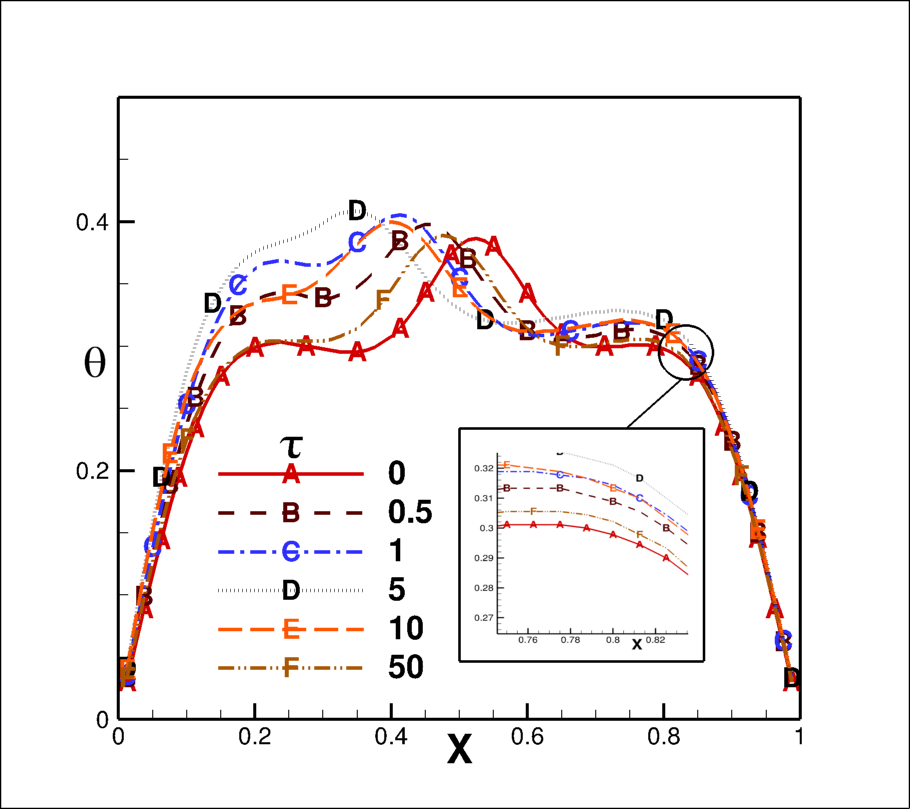}
    \caption{}
    \label{C_CR_Temp}
  \end{subfigure}
  \begin{subfigure}{14cm}
    \centering\includegraphics[width=6cm]{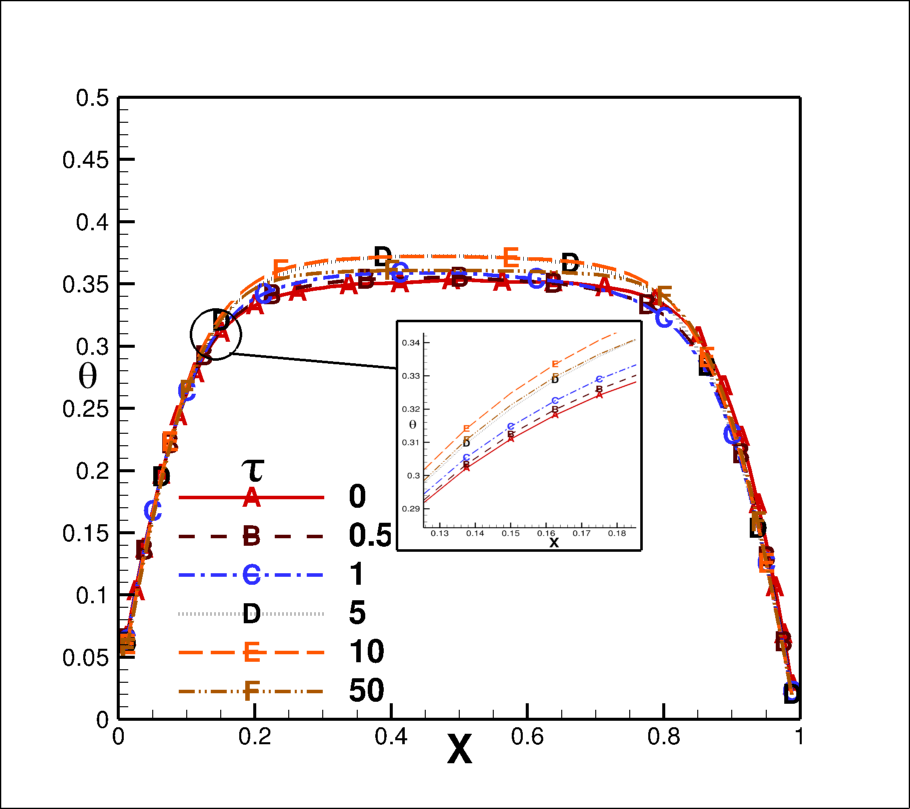}
    \caption{}
    \label{C_CR_top_Temp}
  \end{subfigure}
  \caption{Variation of non-dimensional temperature in horizontal direction at (a) bottom wall (b) mid-height and (c) top wall for different optical thicknesses}
\label{C_CR_BMT}
\end{figure}

The non-dimensional temperature variation in the horizontal direction on the bottom wall, at mid height and top wall are depicted in Fig. \ref{C_CR_BMT} for various optical thicknesses. The non-dimensional temperature curve for the bottom wall has two maxima for optical thickness upto 1 (see Fig. \ref{C_CR_BT}), that corresponds to strike point of the collimated beam and stagnation point developed at the junction of two vortices. Nevertheless, the location of maxima corresponds to strike point of collimated beam remain fixed, it is the highest for radiatively transparent fluid case and keeps on decreasing with increase of optical thickness of fluid. The hot spot due to beam strike does not occur for the medium of optical thickness higher than 5 as the radiative energy gets absorbed with in the fluid before striking on the bottom wall (Fig. 10). The location of second maxima is decided by the relative size of two vortices. The right vortex is smaller to the left vortex for the transparent case, thus stagnation point shifts to the right of the mid point for non-zero optical thickness. The second maximum point is at non-dimensional distance of 0.55, 0.475, 0.4, 0.425, and 0.43 for optical thicknesses 0, 0.5, 1, 5, 10 and 50, respectively. One point to notice is that the highest value of non-dimensional temperature is greater for the transparent case at the location of beam strikes. The second maxima on the non-dimensional temperature curve at the mid height is only achieved on the mid height appear at the location of collimated beam (Fig. \ref{C_CR_Temp}) i.e at left of the vertical at the middle of cavity. The non-dimensional temperature increases on the left part of the curve before second maxima arrives for optical thickness upto 5, afterwards it decreases. Then non-dimensional temperature curve remains almost similar for all optical thickness. There is no maxima in the curve for non-dimensional temperature curve for top wall. The non-dimensional temperature increases from both ends upto distance of 0.21 and remain flat afterwards. The maximum non-dimensional temperature increased to 0.37 for optical thickness 5 and little lesser for other optical thicknesses on the top wall.

\subsubsection{Nusselt number}

\begin{figure}[t]
 \begin{subfigure}{6cm}
   \includegraphics[width=6cm]{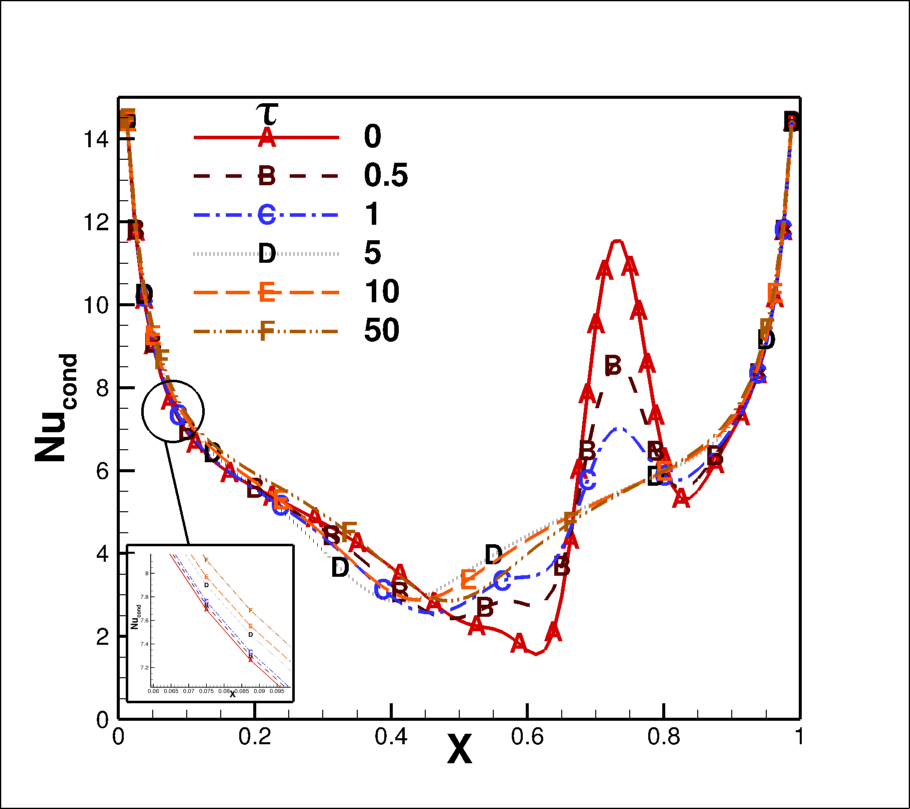}
    \caption{}
    \label{cond_nu}
  \end{subfigure}
   \begin{subfigure}{5cm}
    \includegraphics[width=6cm]{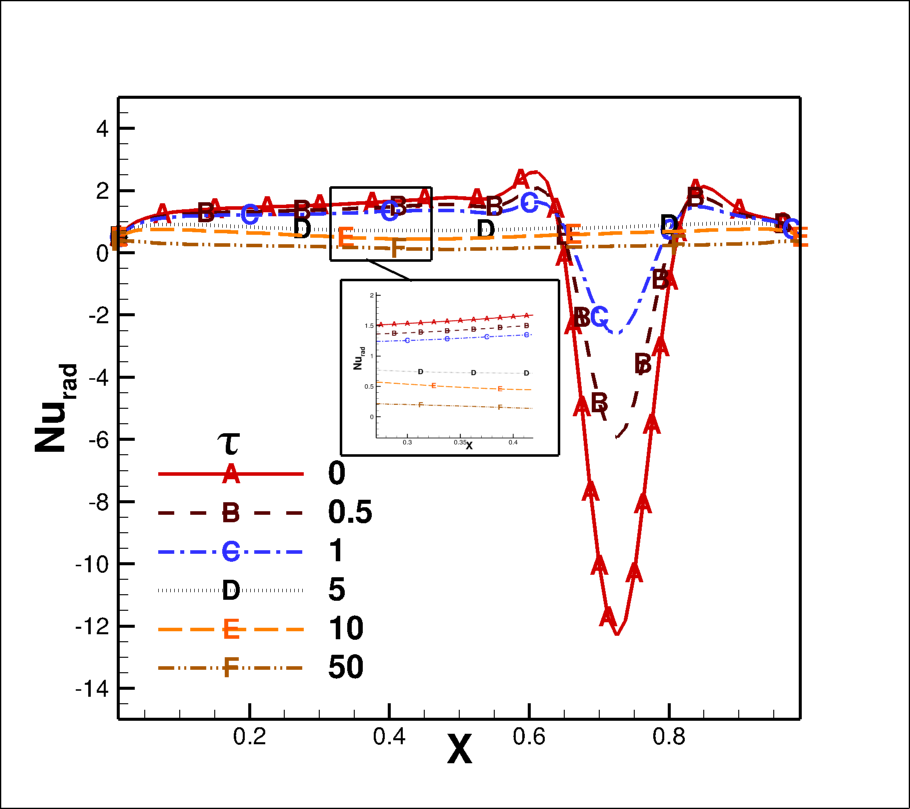}
    \caption{}
    \label{rad_nu}
  \end{subfigure}
  \begin{subfigure}{14cm}
    \centering\includegraphics[width=6cm]{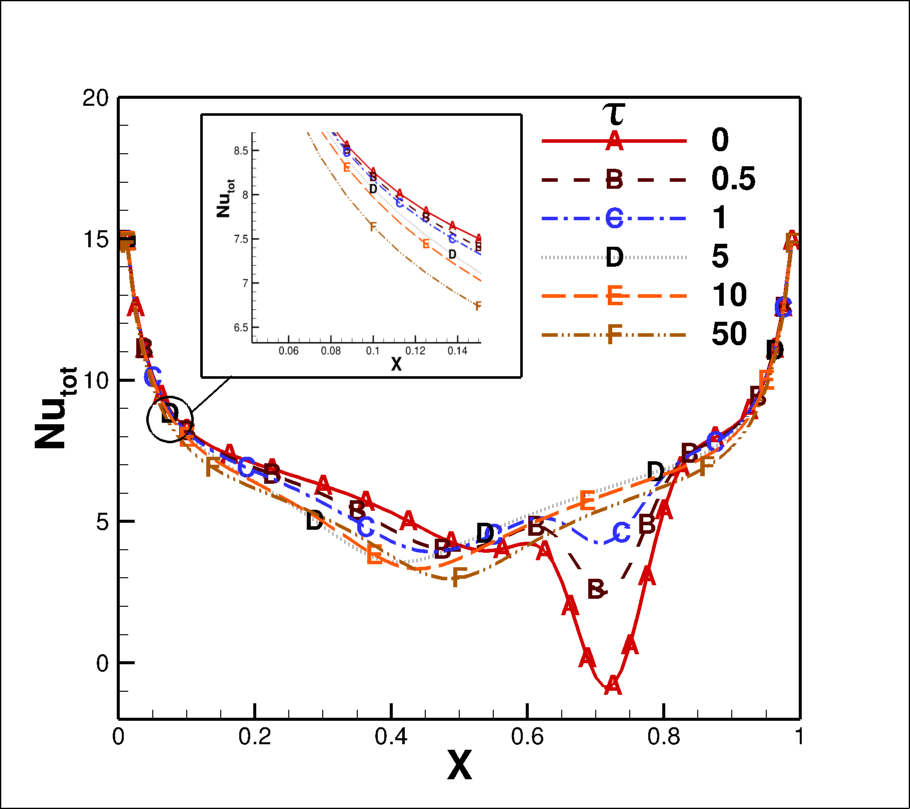}
    \caption{}
    \label{tot_nu}
  \end{subfigure}
    \caption{Variation of (a) conduction (b) radiation and (c) total Nusselt number on bottom wall }
\label{Nu_bot_col}
\end{figure} 

The conduction Nusselt number curve on the bottom wall (Fig. \ref{cond_nu}) decreases drastically upto non-dimensional distance 0.1 from the left corner, afterwards the curve changes its slope and keeps on decreasing till lowest point in the curve. This nature of curve remains same for all optical thicknesses of the fluid, afterwords it shows the effect of collimated beam strike. The conduction Nusselt number curve obtains a minimum value of almost 2 at non-dimensional distance of 0.65 from left corner and all of sudden increases to 12 at the strike length of the collimated beam for transparent case. Further, it decreases to value 6 and then starts increasing and reaches to maximum value of 14 on the right side of the isothermal wall for optical thickness zero case. The lowest value for conduction Nusselt number is obtained at distance of 0.45 and remains constant upto the strike point of beam for optical thickness 0.5, then it reaches to value 1, further its behaviour is similar to the transparent case. Furthermore, the minimum value for conduction Nusselt number obtained at same point and then starts increasing for optical thickness 1 and sudden increase to a value of 7 at strike point of beam. Whereas, the Nusselt number curve behaviour remains unaffected by the collimated beam for optical thickness of fluid 5 and above. The radiation Nusselt number curve is almost constant over whole length of the bottom wall expect the length over which collimated beam strikes. The maximum Nusselt number is 12 which is negative indicates that the radiation energy leaves through radiation mode of heat transfer over this length for transparent case. The Nusselt number curve achieves peak at beam strike length and its value decreases with increase of optical thicknesses of the fluid, also no peak appears for optical thicknesses 10 and 50. One interesting thing is to notice that radiative Nusselt number is almost zero for optical thickness 10 and 50. 

The total Nusselt number which is a linear combination of conduction and radiation Nusselt number is dominated by conduction Nusselt number in the most portion of the length except the length over which collimated beam strikes. The total Nusselt number at the beam strike portion is dominated by the radiation Nusselt number and this portion exhibit the minimum Nusselt number in the curve. The minimum Nusselt number is almost zero which represents the adiabatic condition at the strike point of the collimated beam for the radiation transparent fluid case and this increases to 3 and 5 for optical thickness 0.5 and 1, respectively, whereas no peak appears in the total Nusselt number curve for optical thickness 5, 10 and 50. This is true as collimated energy gets absorbed within the fluid before reaching to the bottom wall.

\begin{figure}[!htb]
 \begin{subfigure}{6cm}
    \includegraphics[width=6cm]{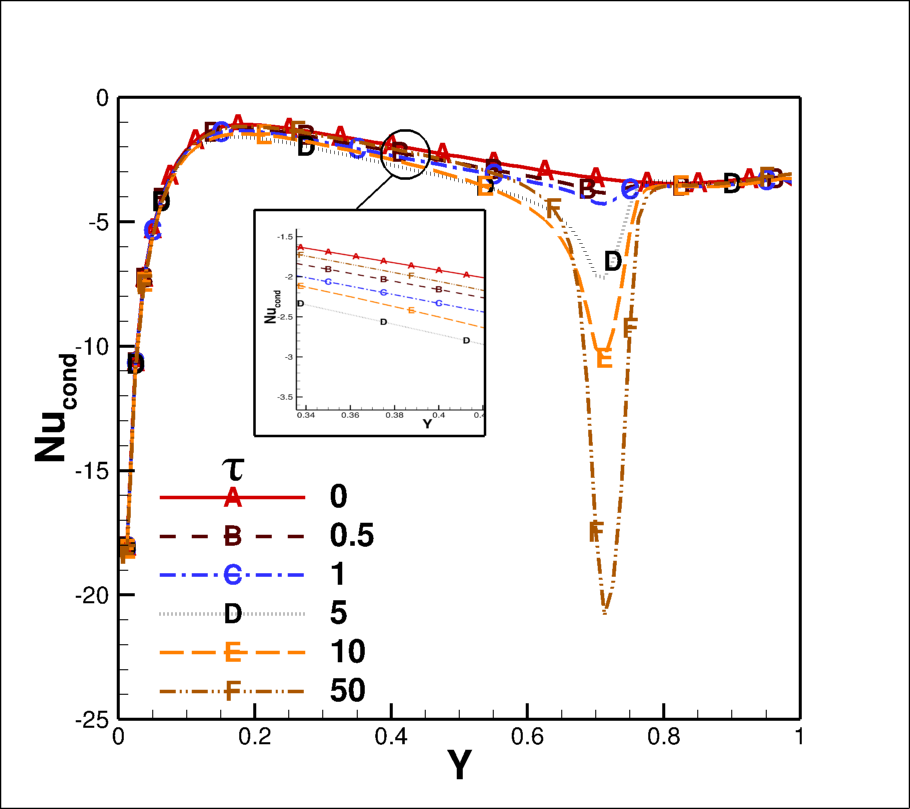}
    \caption{}
    \label{cond_nu_left}
  \end{subfigure}
   \begin{subfigure}{5cm}
    \includegraphics[width=6cm]{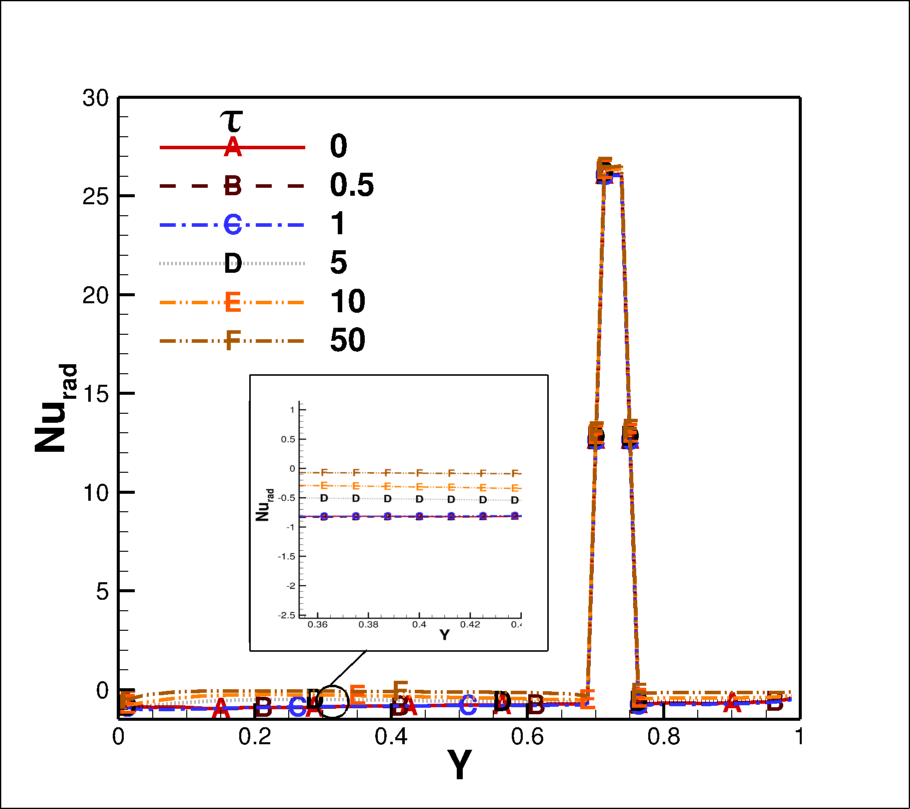}
    \caption{}
    \label{rad_nu_left}
  \end{subfigure}
  \begin{subfigure}{14cm}
   \centering\includegraphics[width=6cm]{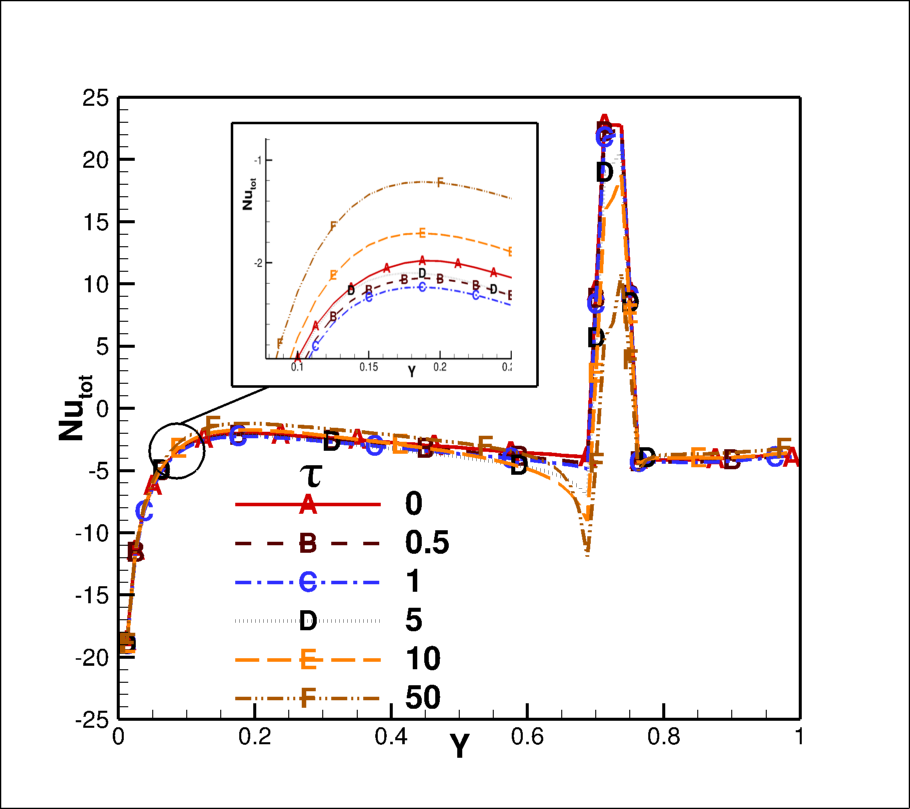}
    \caption{}
    \label{tot_nu_left}
  \end{subfigure}
    \caption{Variation of (a) conduction (b) radiation and (c) total Nusselt number on left wall}
\label{Nu_left_col}
\end{figure}

\begin{figure}[!htb]
    \centering\includegraphics[width=7cm]{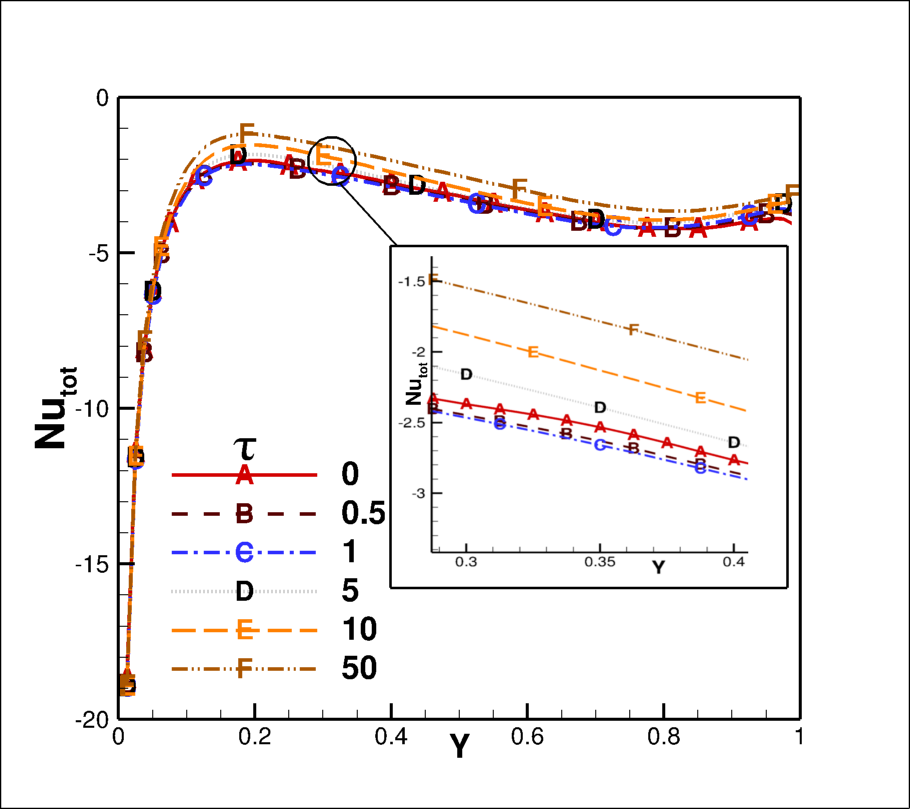}
    \caption{Variation of total Nusselt number on the right wall for various optical thicknesses}
    \label{Nu_right_col}
\end{figure}

The conduction, radiation and total Nusselt number variations on the left wall which also contains the semitransparent window are depicted in Fig. \ref{Nu_left_col} (a), (b), and (c), respectively. The conduction Nusselt number sudden decreases to minimum value at height 0.15 and further, small variation happen upto semitransparent window. There is sudden increase in the conduction Nusselt number on the semitransparent window, this is mostly negative, reveals the that energy leaves from this wall through conduction mode of heat transfer. The conduction Nusselt number value is 20 on the semitransparent window on left wall for the optical thickness 50 case, and this reduces to 3 after semitransparent window and remains constant over rest height of the vertical wall. With the increase in the optical thicknesses of the fluid, the conduction Nusselt number decreases on the semitransparent window, however, optical thicknesses does not have much effect on other portion of the left wall. There is no increase in the conduction Nusselt number at the window for optical thickness 0.5 or optically transparent fluid. The radiative Nusselt number is also negative but very small in number all over the length, except on the semitransparent window (Fig. \ref{rad_nu_left}), where radiative Nusselt number is positive, reveals the radiative flux is coming inside the cavity through the window. There is mimimum variation of radiation Nusselt number with optical thickness over the length except at the window, where Nusselt number is 26 and remains same for all optical thicknesses. The total Nusselt number curve which is linear combination of conduction and radiation Nusselt number is shown in Fig. \ref{tot_nu_left}. The total Nusselt number is dominated by conduction over all height of the wall except around the window where radiative Nusselt number is in dominance. The total Nusselt number is negative indicates that the energy is going out from the isothermal wall except the window where Nusselt number is positive which indicates that the energy is coming inside from the window. Figure. \ref{Nu_right_col}, depicts the variation of total Nusselt number on the right wall. This wall does not have any semitransparent window, thus the total Nusselt number behaviour is similar to the left wall without any phenomenon  that happens on the semitransparent window. The total Nusselt number is negative reveals that the heat is also being transferred outside from the right side wall and most heat is transferred within few small height of the wall afterward, heat transfer rate is small. 

\begin{landscape}
\begin{table}
\centering
\caption{Average Nusselt number values on the different walls for various value of optical thicknesses of fluid}
\label{co_avgNu_table}
\begin{tabular}{|cccccccccc|}
\hline
\multirow{3}{*}{Optical thickness} & \multicolumn{9}{c|}{Average Nusselt number} \\ \cline{2-10} 
 & \multicolumn{3}{c}{Bottom wall} & \multicolumn{3}{c}{Left wall} & \multicolumn{3}{c|}{Right Wall} \\ \cline{2-10} 
 & Conduction & Radiation & Total & Conduction & Radiation & Total & Conduction & Radiation & Total \\ \hline
0 & 5.84 & 0.154 & 5.994 & -2.845 & 0.58 & -2.264 & -2.988 & -0.778 & -3.766 \\ \hline
0.5 & 5.662 & 0.642 & 6.305 & -3.042 & 0.55 & -2.491 & -3.021 & -0.792 & -3.813 \\ \hline
1 & 5.602 & 0.848 & 6.45 & -3.168 & 0.546 & -2.621 & -3.052 & -0.777 & -3.829 \\ \hline
5 & 5.661 & 0.8 & 6.462 & -3.602 & 0.779 & -2.823 & -3.137 & -0.509 & -3.646 \\ \hline
10 & 5.724 & 0.601 & 6.325 & -3.817 & 0.961 & -2.855 & -3.127 & -0.343 & -3.47 \\ \hline
50 & 5.812 & 0.224 & 6.036 & -4.099 & 1.208 & -2.89 & -3.037 & -0.109 & -3.146 \\ \hline
\end{tabular}
\end{table}
\end{landscape}

Table \ref{co_avgNu_table} shows the average Nusselt number on the different walls of the cavity. The average conduction Nusselt number on the bottom wall decreases slowly with optical thickness upto 1 then starts increasing whereas treand for average radiation Nusselt number is faster and reverse to the conduction Nusselt number making total Nusselt number increase upto optical thickness 5, then decreases on the bottom wall. The average conduction and radiation Nusselt numbers increase for all optical thicknesses on the left wall but in reverse direction, i.e, conduction Nusselt number is negative and radiative Nusselt number is positive. The rate of increase of both the Nusselt numbers is such that total Nusselt number increases with increase of optical thickness of the fluid on the left wall. Furthermore, the conduction Nusselt number increases upto optical thickness 5 then decreases, whereas radiative Nusselt number only increases upto optical thickness 1 then decreases on the right wall. The linear combination of these two Nusselt numbers make total Nusselt number, that increase upto optical thickness 5 then decreases. All the Nusselt number on the right side wall is negative indicates that energy goes out by both modes of heat transfer through this wall.

\section{Conclusions}
The effect of diffuse and collimated beam radiation on the symmetrical cooling case of natural convection in a cavity heated from bottom has been thoroughly investigated. The collimated beam radiation feature have been developed in OpenFOAM framework and an applications has been made by including libraries of other fluid flow and heat transfer models present in OpenFOAM package. The collimated beam case has been simulated by creating semitransparent window of non-dimensional width 0.05 at non-dimensional height of 0.7 on the left wall and a collimated beam of value 1000 $W/m^2$ is irradiated on semitransparent window at an angle of $45^0$. The following conclusions are drawn from the comprehensive study of fluid flow and heat transfer phenomena in the scenarios of diffuse and collimated beam radiation for various optical thickness of fluid inside the cavity.

\begin{enumerate}
\item The isotherm and stream lines are symmetrical about the vertical lines at the middle of the cavity and formation of two counter rotating vortices happen inside the cavity for diffuse radiation case.
\item The maximum values of non-dimensional temperature increases with the increase of optical thickness of the fluid whereas, maximum value of non-dimensional stream function increases upto optical thickness 5 then remains almost constant.
\item These symmetrical vortices change its dynamics by collimated incidence for different values of optical thickness of the fluid. The right side vortex is smaller than the left side for transparent fluid, whereas, left side vortex is smaller for non-zero optical thickness of the fluid.
\item The isotherms are slightly bent in the path of collimated beam for non-zero optical thicknesses of fluid.
\item The variation of Nusselt numbers are mostly dominated by conduction Nusselt number over the length of the walls except at the portion of the collimated irradiation, where radiative Nusselt number is dominating for lower value of optical thickness of fluid.
\item The average total Nusselt number increases with increase of optical thickness upto 5 on the bottom as well asright wall and it increases for all optical thicknesses on the left wall.
\item The energy enters into the cavity by conduction and radiative mode of heat transfer through bottom wall, whereas energy enters through radiation modes of heat transfer and leave by conduction mode of heat transfer from the left wall while energy leaves through both mode of heat transfer from the right wall.

\end{enumerate}
\section*{Acknowledgements}
The authors greatly acknowledge the financial support provided by Science and Engineering Research Board (SERB) (Statutory Body of the Government of India) via Grant.No:ECR/2015/000327 to carry out the present work.

\section{Declaration of interests}
The authors declare that they have no known financial interests or personal relationships that could have appeared to influence the work reported in this paper.

  \nomenclature[a]{$a$}{Co-efficient}
    \nomenclature[a]{$C_{p}$}{Specific heat capacity ($J/kg-K$)}
    \nomenclature[a]{$g$}{Acceleration due to gravity ($m/s^2$)}
    \nomenclature[a]{$G$}{Irradiation ($W/m^2$)}
    \nomenclature[a]{$H$}{Height ($m$)}
    \nomenclature[a]{$I$}{Intensity ($W/m^2$)}
    \nomenclature[a]{$I_{b}$}{Black body intensity ($W/m^2$)}
    \nomenclature[a]{$k$}{Thermal conductivity ($W/mK$)}
    \nomenclature[a]{$L$}{Length of the domain of study ($m$)}
    \nomenclature[a]{$Nu$}{Nusselt number}
    \nomenclature[a]{$p$}{Pressure ($N/m^2$)}
    \nomenclature[a]{$Pr$}{Prandtl number}
	\nomenclature[a]{$q$}{Flux ($W/m^2$)}
	\nomenclature[a]{$Ra$}{Rayleigh number}
	\nomenclature[a]{$u,v$}{Velocity ($m$)}
    \nomenclature[g]{$\beta_{T}$}{Thermal expansion coefficient ($1/K$)}
    \nomenclature[g]{$\epsilon$}{Emissivity}
	\nomenclature[g]{$\kappa_{a}$}{Absorption coefficient ($1/m$)} 
	\nomenclature[g]{$\rho$}{Density of the fluid ($kg/m^3$)}
    \nomenclature[g]{$\tau$}{Optical thickness}
	\nomenclature[g]{$\phi$}{scalar}
    \nomenclature[U]{$C$}{Conduction}	
	\nomenclature[U]{$c$}{Cold wall}
	\nomenclature[U]{$co$}{collimated beam}
	\nomenclature[U]{$conv$}{Convection}
	\nomenclature[U]{$f$}{Face centre}
	\nomenclature[U]{$free$}{Free stream}
	\nomenclature[U]{$i,j$}{Tensor indices}
	\nomenclature[U]{$nb$}{Neighbour cell}
	\nomenclature[U]{$p$}{Cell centre}		
	\nomenclature[U]{$R$}{Radiation}
	\nomenclature[U]{$ref$}{Reference}
	\nomenclature[U]{$t$}{Total}
	\nomenclature[U]{$w$}{Wall}
    
\printnomenclature




\bibliography{mybibfile}

\begin{thebibliography}{10}
\expandafter\ifx\csname url\endcsname\relax
  \def\url#1{\texttt{#1}}\fi
\expandafter\ifx\csname urlprefix\endcsname\relax\def\urlprefix{URL }\fi
\expandafter\ifx\csname href\endcsname\relax
  \def\href#1#2{#2} \def\path#1{#1}\fi

\bibitem{Torrance}
{K.E. Torrance, J.A. Rockett}, Numerical study of natural convection in an
  enclosure with localized heating from below-creeping flow to the onset of
  laminar instability, J.Fluid Mech, part 1 36 (1969) 33--54.
\newblock \href {https://doi.org/10.1017/S0022112069001492}
  {\path{doi:10.1017/S0022112069001492}}.

\bibitem{Calcagani}
{B. Calcagani, F. Marsili, M. Paroncini}, Natural convective heat transfer in
  square enclosures heated from below, App. Thermal engineering 25 (2005)
  2522--2531.
\newblock \href {https://doi.org/10.1016/j.applthermaleng.2004.11.032}
  {\path{doi:10.1016/j.applthermaleng.2004.11.032}}.

\bibitem{Ganzorolli}
{M.M. Ganzorolli, L.F. Milanez}, Natural convection in rectangular enclosures
  heated from below and symmetrically cooled from the sides, Int.J. Heat mass
  Transfer, (6) 36 (1995) 1063--1073.
\newblock \href {https://doi.org/10.1016/0017-9310(94)00217-J}
  {\path{doi:10.1016/0017-9310(94)00217-J}}.

\bibitem{Aydin}
{Orhan Aydin, Wen-Hei Yang}, Natural convection in enclosures with localized
  heating from below and symmetrical cooling from sides, Int. J of Num. Methods
  for Heat \& Fluid flow, No.5 10 (2000) 518--529.
\newblock \href {https://doi.org/10.1108/09615530010338196}
  {\path{doi:10.1108/09615530010338196}}.

\bibitem{Acharya}
{S. Acharya, R.J. Goldstein}, Natural convection in an externally heated
  vertical or inclined square box containing internal energy sources, J. Of
  Heat Transfer 107 (1985) 855--866.
\newblock \href {https://doi.org/10.1115/1.3247514}
  {\path{doi:10.1115/1.3247514}}.

\bibitem{Webb}
{B. W. Webb, R. Viskanta}, Radiation-induced buoyancy driven flow in
  rectangular enclosures: Experiment and analysis, J.of Heat Transfer 109
  (1987) 427--433.
\newblock \href {https://doi.org/10.1115/1.3248099}
  {\path{doi:10.1115/1.3248099}}.

\bibitem{Mezrhab}
{A. Mezrhab, H.Bouali, H. Amaoui, M. Bouzidi}, Compuations of combined natural
  convection and radiation heat transfer in a cavity having a square body at
  its centre, App.Energy 83 (2006) 1004--1023.
\newblock \href {https://doi.org/10.1016/j.apenergy.2005.09.006}
  {\path{doi:10.1016/j.apenergy.2005.09.006}}.

\bibitem{Hua}
{Hua Sun, Eric Chenier, Guy Lauriat}, Effect of surface radiation on the
  breakdown of steady natural convection flows in a square, air-filled cavity
  containing a centred inner body, App. Thermal Engineering 31 (2011)
  252--1262.
\newblock \href {https://doi.org/10.1016/j.applthermaleng.2010.12.028}
  {\path{doi:10.1016/j.applthermaleng.2010.12.028}}.

\bibitem{Mukul}
{Mukul Paramanda, Salman Khan, Amaresh Dalal, Ganesh Natarajan}, Critical
  assessment of numerical alogorithms for convective-radiative heat transfer in
  enclosures with different geometries, Int.J. of Heat and Mass Transfer 108
  (2017) 627--644.
\newblock \href {https://doi.org/10.1016/j.ijheatmasstransfer.2016.12.033}
  {\path{doi:10.1016/j.ijheatmasstransfer.2016.12.033}}.

\bibitem{Kumar}
{P. Kumar, V. Eswaran}, The effect of radiation on natural convection in
  slanted cavities of angle $45^0$ and $60^0$, Int.J.Thermal Science 67 (2013)
  96--106.
\newblock \href {https://doi.org/10.1016/j.ijthermalsci.2012.12.009}
  {\path{doi:10.1016/j.ijthermalsci.2012.12.009}}.

\bibitem{Saravanan}
{S. Saravanan, C. Sivaraj}, Coupled thermal radiation and natural convection
  heat transfer in a cavity with a heated plate inside, Int. J. Heat and Fluid
  Flow 40 (2013) (2013) 54--64.
\newblock \href {https://doi.org/10.1016/j.ijheatfluidflow.2013.01.007}
  {\path{doi:10.1016/j.ijheatfluidflow.2013.01.007}}.

\bibitem{Bittagopal}
{B. Mondal, S. C. Mishra}, Simulation of natural convection in the presence of
  volumetric radiation using the lattice boltzmann method, Num.Heat
  Transfer,part-A 55 (2009) 18--41.
\newblock \href {https://doi.org/10.1080/10407780802603121}
  {\path{doi:10.1080/10407780802603121}}.

\bibitem{Yujia}
{Yujia Sun, Xiaobing Zhang, J. R. Howell}, Assessment of different radiative
  transfer equation solvers for combined natural convection and radiation heat
  transfer problems, J. of Quantitative Spectroscopy and Radiative Transfer 194
  (2017) 31--46.
\newblock \href {https://doi.org/10.1016/j.jqsrt.2017.03.022}
  {\path{doi:10.1016/j.jqsrt.2017.03.022}}.

\bibitem{Yuan}
{Xing Yuan, Fatemeh Tavakkoli, Kambiz Vafai}, Analysis of natural convection in
  horizontal concentric annuli of varying inner shape, Numerical Heat Transfer,
  Part A 68 (2015) 1155–1174.
\newblock \href {https://doi.org/10.1080/10407782.2015.1032016}
  {\path{doi:10.1080/10407782.2015.1032016}}.

\bibitem{Anand}
{N. Anand Krishna, S. C. Mishra}, Discrete transfer method applied to radiative
  transfer in variable refractive index, J.of Quantative Spectroscopy and
  Radiative Transfer 102 (2006) 432--440.
\newblock \href {https://doi.org/10.1016/j.jqsrt.2006.02.024}
  {\path{doi:10.1016/j.jqsrt.2006.02.024}}.

\bibitem{Ben}
{P. Ben Abdallah, V. Le Dez}, Radiative flux field inside an absorbing-emitting
  semi transparent slab with a variable refractive index at radiative
  conductive coupling, J.of Quantative Spectroscopy and Radiative Transfer
  {67(2)} (2000) 125--137.
\newblock \href {https://doi.org/10.1016/S0022-4073(99)00200-9}
  {\path{doi:10.1016/S0022-4073(99)00200-9}}.

\bibitem{Ya}
{Ya. A. Ilyushin}, Propagating of collimated beam in the refractive scattering
  medium, Radiophysics and Quantum Electronics 55 (2013) 10–11.
\newblock \href {https://doi.org/10.1007/s11141-013-9402-8}
  {\path{doi:10.1007/s11141-013-9402-8}}.

\bibitem{Anil}
{Anil K. Verma, P. Rath, S. K. Mahapatra}, Interaction of short pulse
  collimated irradiation with inhomogeneity: An accurate model, Int. Com. in
  Heat and Mass Transfer 72 (2016) 1–9.
\newblock \href {https://doi.org/10.1016/j.icheatmasstransfer.2015.08.025}
  {\path{doi:10.1016/j.icheatmasstransfer.2015.08.025}}.

\bibitem{Rath}
{P. Rath, S. K. Mahapatra}, New formulation of radiative flux in ultrashort
  time scale with its implications, J. of Thermophysics and Heat transfer, No 2
  26 (2012) 294--299.
\newblock \href {https://doi.org/10.2514/1.T3793} {\path{doi:10.2514/1.T3793}}.

\bibitem{openfoam2017open}
OpenFOAM, The open source cfd toolbox: User guide, openfoam v1706 (2017).

\bibitem{patankar}
{S. V. Patankar}, Numerical heat transfer and fluid flow, Hemisphere Publishing
  Corporation, 1980.

\bibitem{Moukalled}
{F. Moukalled, L. Mangani, M. Darwish}, The Finite Volume Method in
  Computational Fluid Dynamics: An Advanced Introduction with OpenFOAM and
  Matlab, Springer International Publishing, 2016.

\bibitem{RAD19}
{Ankur Garg, G Chanakya, Pradeep Kumar}, Numerical error estimation in finite
  volume method for radiative transfer equation for collimated irradiation, in:
  Proceedings of the 9th International Symposium on Radiative Transfer, RAD-19,
  June 3-7, 2019, Athens, Greece, 2019.

\bibitem{Aswatha}
{Aswatha, C. J. Gangadhara Gowda, S. N. Sridhara, K. N. Seetharamu}, Effect of
  different thermal boundary conditions at bottom wall on natural convection in
  cavities, J. of Engineering Science and Technology 6 (2011) 109 -- 130.

\bibitem{Lari}
{K. Lari, M. Baneshi, S. G. Nassab, A. Komiya, S. Maruyama}, Combined heat
  transfer of radiation and natural convection in a square cavity containing
  participating gases, Int.J.Heat Mass Transfer 54 (2011) 5087--5099.
\newblock \href {https://doi.org/10.1016/j.ijheatmasstransfer.2011.07.026}
  {\path{doi:10.1016/j.ijheatmasstransfer.2011.07.026}}.

\end{thebibliography}

\end{document}